\documentclass[10pt,aps,pra,letterpaper,twocolumn,superscriptaddress,floatfix,nofootinbib,longbibliography]{revtex4-2}
\usepackage[utf8]{inputenc}
\usepackage{graphicx}
\usepackage{amsmath,amssymb,amsfonts,amsthm}
\usepackage{mathtools}
\usepackage{mathrsfs}
\usepackage{bm}
\usepackage{physics}
\usepackage{array}
\usepackage{multirow}
\usepackage{booktabs}
\usepackage{blkarray}
\usepackage{xcolor}
\usepackage{pgf,tikz}
\usepackage{hyperref}
\hypersetup{colorlinks=true,linkcolor=blue,citecolor=blue,urlcolor=blue}
\graphicspath{{pict/}{}}

\newcommand{\Cov}{\mathrm{Cov}}

\makeatletter
\let\SM@hangfrom@section\@hangfrom@section
\makeatother

\begin{document}

% the table of contents of the Supplementary Material lists only its own sections
\let\SIaddcontentsline\addcontentsline
\renewcommand{\addcontentsline}[3]{}

\title{Optimal pure state estimation with adaptive Fisher-symmetric measurements}

\author{C. Vargas}
\email{covargas2019@udec.cl}
\affiliation{Instituto Milenio de Investigaci\'on en \'Optica, Universidad de Concepci\'on, Concepci\'on, Chile}
\affiliation{Facultad de Ciencias F\'isicas y Matem\'aticas, Departamento de F\'isica, Universidad de Concepci\'on, Concepci\'on, Chile}
\author{A. Delgado}
\email{aldelgado@udec.cl}
\affiliation{Instituto Milenio de Investigaci\'on en \'Optica, Universidad de Concepci\'on, Concepci\'on, Chile}
\affiliation{Facultad de Ciencias F\'isicas y Matem\'aticas, Departamento de F\'isica, Universidad de Concepci\'on, Concepci\'on, Chile}
\author{L. Pereira}
\email{luciano.pereira@icfo.eu}
\affiliation{ICFO - Institut de Ciencies Fotoniques, The Barcelona Institute of Science and Technology, 08860 Castelldefels, Barcelona, Spain}

\begin{abstract}
Quantum state estimation is important for various quantum information processes, including quantum communications, computation, and metrology, which require characterizing quantum states for evaluation and optimization. We present a three-stage adaptive estimation method that is asymptotically optimal for estimating arbitrary $d$-dimensional pure quantum states. The protocol uses locally informationally complete Fisher symmetric measurements (FSM) and a measurement on an arbitrary basis. We derive an asymptotic high-probability error bound for the protocol and show that its infidelity approaches the optimal value $(d-1)/N$ given by the Gill-Massar bound (GMB) for sample sizes $N\gg d^2$. This is verified by numerical simulations up to $d=64$. Furthermore, we apply the protocol to weakly mixed states, obtaining an accuracy close to the GMB. The total number of measurement outcomes is $7d-3$, scaling linearly with $d$, avoiding the need for collective measurements on multiple copies of the unknown state.
\end{abstract}

\maketitle

\section{Introduction}\label{sec1}

Quantum state estimation \cite{Helstrom,Holevo,Teo2015} plays an important role in quantum information processes such as quantum communications, quantum computing \cite{Nielsen}, and quantum metrology \cite{Giovannetti,Nagata,Okamoto1,Xiang,Jones}, whose comparative evaluation and optimization require the characterization of quantum states. The design of quantum state estimation methods balances the experimental feasibility and accuracy of the estimation, which are quantified by the total number of measurement outcomes and the infidelity \cite{Jozsa1993,Braunstein,Paris2009}, respectively, as functions of dimension $d$ and sample size $N$ \cite{Straupe2016,Lange2023,Chen2023,Lowe2025,Haah2017}.
 
The accuracy of any estimation method based on separable measurements is fundamentally limited by the Gill-Massar bound (GMB) \cite{GillGuta2013,Munoz2022,Gill}. Achieving this bound is challenging because optimal measurements typically depend on the unknown state. Adaptive tomography methods address this problem by obtaining several estimates sequentially, with each estimate used to select the subsequent measurement \cite{Bagan2002,Bagan2006,Hayashi2008,Mahler2013,Okamoto,Hou2016,pereira2018,Struchalin2018}. For qubits, two-stage adaptive methods have been shown to reach the GMB asymptotically \cite{Gill,Bagan2002,Bagan2006,Hayashi2008}, but for density matrices of dimension $d>2$ the bound cannot be saturated in general \cite{Gill,pereira2018,Struchalin2018,pereira2020,ZhuThesis,Zhu}. In contrast, for pure states of any dimension the GMB is attainable \cite{Li,Zhu}, and two measurement bases adapted to a preliminary estimate reach it asymptotically \cite{Girotti2024}. However, these results provide neither error bounds for a finite sample size nor the rate of convergence to the GMB.
 
Fisher symmetric measurements (FSM) \cite{Li,FSM_EXP,ZhuThesis} are an example of measurements that attain the GMB for pure states. They provide uniform and maximal information about all parameters that characterize a quantum state and, for pure states, are implemented by a positive operator-valued measure (POVM) with at least $2d-1$ rank-1 elements. Thus, FSMs use the smallest possible number of measurement outcomes and achieve the best possible accuracy with separable measurements on a sample of identically prepared copies of the unknown state. However, they depend on the unknown state, that is, they are only locally informationally complete, and the unknown state must be in the neighborhood of a previously known fiducial state. It has been shown \cite{Zhu} that FSMs can be designed without a priori information, but this requires a POVM that acts collectively on two copies of the unknown state, with up to $4d^2$ measurement outcomes. FSMs have also been introduced for mixed states, but are not optimal \cite{Li,Zhu}.
 
In this work, we propose a three-stage adaptive estimation protocol for unknown pure states, based on three FSMs and a measurement on an arbitrary basis. This protocol is asymptotically optimal, that is, it achieves the GMB for large sample sizes, and requires no other a priori information apart from the purity of the unknown state. In this way, we extend the applicability of FSMs to any unknown pure state. We first show that, for a given fiducial state, two FSMs together estimate all pure states except those orthogonal to the fiducial state. To avoid this case, we introduce a preliminary stage in which we measure a small sample of the unknown state on an arbitrary basis and select the most frequent outcome as the fiducial state. Due to this, these FSMs are not necessarily optimal for the unknown state, but they still allow one to obtain a first estimate of the unknown state that is close enough to serve as the fiducial state of a third, near-optimal FSM.
 
We derive high-probability bounds for two state estimators, an analytical linear estimator, and a maximum likelihood estimator (MLE). For the linear estimator, we obtain a conservative finite-sample bound, valid for any sample size $N$, which shows the advantage of adaptation for large $N$. For the MLE, we obtain an asymptotic bound, which improves the bound for the linear estimator and shows that, with an optimal split of the sample size between the stages, the average infidelity approaches the optimal value $(d-1)/N$ given by the GMB for $N\gg d^2$, while a single estimate fluctuates around it by $\mathcal{O}(\sqrt{d}/N)$. Numerical simulations up to $d=64$ confirm this behavior and show that the protocol also provides an accuracy close to the GMB for weakly mixed states. 

The total number of measurement outcomes of our protocol is $7d-3$, which scales linearly with $d$, and the protocol does not rely on collective measurements on multiple copies of the unknown state. Each of the three FSMs can be replaced by two measurement bases, at the expense of increasing the total number of measurement outcomes to $7d$. It has been shown that any pure state can be estimated with five measurement bases \cite{Goyeneche}. Thus, adding two more measurement bases allows us to reach the GMB asymptotically. Collective measurements can also be avoided with fully adaptive estimation methods, in which each measurement is adapted using the results of the previous ones \cite{Okamoto}. In these methods, estimates are obtained from local optimal measurements \cite{Nagaoka1,Nagaoka2,Nagaoka3,Fujiwara1} or from a global estimation strategy \cite{Ferrie,Utreras,Zambrano,granade2017,Farooq2022}. However, these methods require a large number of iterations, which, unlike our protocol, increases the total number of measurement outcomes.

\section{Results}\label{sec2}

\subsection{Fisher symmetric measurements}\label{subsec:fsm}
 
The Gill-Massar bound (GMB) establishes a fundamental limitation on the estimation of a pure state $\ket*{\Psi}$ from a sample of size $N$. Let $I(\Psi,\hat\Psi)=1-|\braket*{\hat\Psi}{\Psi}|^2$ be the infidelity between $\ket*{\Psi}$ and its estimator $\ket*{\hat\Psi}$. According to the GMB, the average infidelity in the set $\Lambda$ of estimates of a locally unbiased estimator is lower bounded by
	\begin{equation}
	   \bar{I}(\Psi) = \int_\Lambda {\rm d}\hat\Psi I(\Psi,\hat\Psi) \geq  \bar{I}_{\rm GM}=\frac{d-1}{N}. \label{eq:GMB}
    \end{equation}
An estimation protocol is optimal when it saturates the GMB. Fisher Symmetric Measurements (FSM) achieve this. Let us consider the $d$-dimensional unknown pure state 
    \begin{align}
        |\Psi\rangle=a_0|0\rangle+\sum_{k=1}^{d-1}z_k|k\rangle,
    \end{align}
with $z_k\in\mathbb{C}$, $|z_k|^2\ll1$, and $a_0^2=1-\sum_{k=1}^{d-1}|z_k|^2\approx 1$, so that it lies in the neighborhood of the state $|0\rangle$, and an arbitrary POVM of $n$ rank-1 operators $\mathcal{E}=\{|\varphi^\alpha\rangle\langle\varphi^\alpha|\}_{\alpha=1}^n$, with $|\varphi^\alpha\rangle=\beta_0^\alpha|0\rangle+\sum_{k=1}^{d-1}\omega_k^\alpha|k\rangle$, $\beta_0^\alpha> 0$ and $\omega_k^\alpha\in\mathbb{C}$ for $k\in\{1,\dots,d-1\}$ \cite{Yuen, Liu2020}. POVM $\mathcal{E}$ is an FSM for state $\ket*{\Psi}$ if it saturates the GMB of Eq.\thinspace\eqref{eq:GMB} and if all parameters $\{z_k\}$ are determined with the same precision. Both conditions are fulfilled when $\sum_{\alpha=1}^n\omega_j^\alpha \omega_k^\alpha=0$ for $j,k\in\{1,\dots,d-1\}$ and $n\geq 2d-1$. According to this, FSMs achieve optimal accuracy in estimating pure states close to $\ket*{0}$, that is, $\bar{I}(\Psi)\approx \bar{I}_{\rm GM}$, with a minimal number of outcomes, $n=2d-1$.
 
FSMs present two drawbacks. First, there is no analytical inversion method to derive an estimator of $\ket*{\Psi}$ from the FSM $\mathcal{E}$, so a statistical inference tool must be used, such as maximum likelihood estimation (MLE). Second, FSMs are locally informationally complete, that is, they can only accurately reconstruct states close to $|0\rangle$. It remains unclear how close a state $\ket*{\Psi}$ needs to be to $\ket*{0}$ to be effectively estimated. In the Supplementary Material (SM), we address both drawbacks. First, we present an analytical procedure to obtain an FSM estimator from measurement statistics (SM, Sect. \ II \, D). Second, we show that this estimator provides an infidelity of order $d/(a_0^2N)$, while the linear approximation introduces a bias that fundamentally limits the accuracy of the estimation (SM, Sect. \ II \, E),
    \begin{align}
         I(\Psi,\hat\Psi) \lesssim \frac{4\kappa^2}{a_0^2}, \quad \kappa = (1-a_0^2)\sqrt{\frac{2d-1}{2}},
    \end{align}    
which is independent of $N$. We can see that the approximation error $\kappa$ is small when $a_0^2\gg1-\sqrt{2/(2d-1)}$, that is, the infidelity $1-a_0^2$ between $\ket*{\Psi}$ and $\ket*{0}$ has to decrease as $\mathcal{O}(\sqrt{1/d})$. However, to obtain an infidelity close to the GMB, the approximation error must also be smaller than the statistical error.
 
\subsection{Adaptive Fisher symmetric measurements}\label{subsec:afsm}
 
\begin{figure*}[t]\centering
		\includegraphics[width=0.7\linewidth]{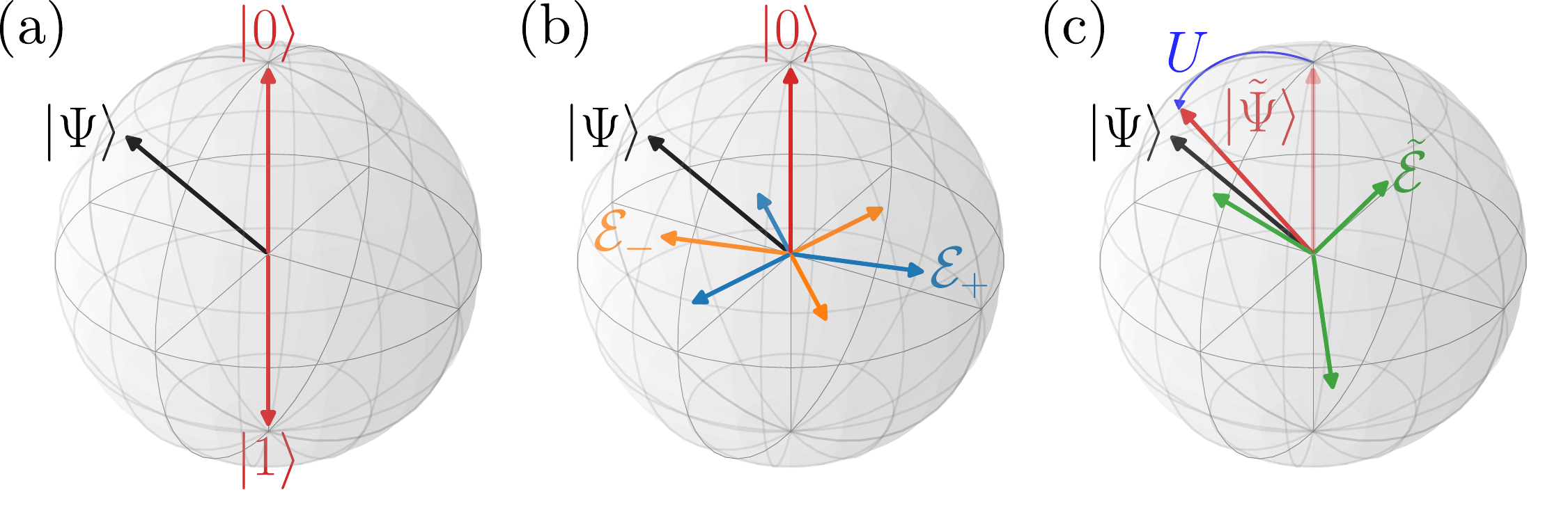}
		\caption{Schematic representation of the adaptive Fisher-Symmetric measurements for an unknown pure state $|\Psi\rangle$ (black). (a) Measurement in the basis $\{|0\rangle,|1\rangle\}$ (red) for selecting a fiducial state. (b) Two-FSMs $\mathcal{E}_{\pm}$ (blue and orange) with fiducial state $|0\rangle$ (red). (c) Adapted FSM $\tilde{\mathcal{E}}$ (green) with fiducial state $|\tilde{\Psi}\rangle$ (red) estimated in the previous step. The unitary $U$ (blue) is a change of basis such that $|\tilde{\Psi}\rangle = U|0\rangle$. }
		\label{Figure1}
	\end{figure*}
 
We consider two FSMs, shown in Fig.\thinspace\ref{Figure1}(b), given by POVMs $\mathcal{E}_\pm = \{|\varphi^\alpha_\pm\rangle\langle\varphi^\alpha_\pm|\}_{\alpha=1}^n$ with
	\begin{equation}
		|\varphi^\alpha_\pm\rangle=\beta_0^\alpha|0\rangle\pm\sum_{k=1}^{d-1}\omega_k^\alpha|k\rangle,
	\end{equation}
where the coefficients $\{\omega_k^\alpha\}$ satisfy the conditions of being an FSM. Each of these FSMs estimates pure states in a neighborhood of the fiducial state $|0\rangle$. However, in the SM, Sec.\ III, we show that, taken together, they can estimate an arbitrary pure state $|\Psi\rangle$ except those orthogonal to the fiducial one. In fact, from the measurement statistic $P_\pm^\alpha=|\langle\varphi_\pm^\alpha|\Psi\rangle|^2$ we find that
    \begin{align}
		a_0 z_k &= \frac{1}{2}\sum_{\alpha=1}^n \left(\frac{\omega_k^\alpha}{\beta_0^\alpha}\right)(P_+^\alpha -P_-^\alpha), \label{system_eq}
    \end{align}
which together with the normalization condition $\sum_{k=1}^{d-1}|z_k|^2=1 - a_0^2$ allows us to directly determine a first estimate $|\tilde{\Psi}\rangle$ of the state $|\Psi\rangle$, except when $|\Psi\rangle$ has no overlap with the fiducial state $|0\rangle$. This case can be avoided by measuring a sample of size $N_0\approx 10d$ on an arbitrary basis, as shown in Fig.\thinspace\ref{Figure1}(a). Since the largest probability satisfies $p_{\max}\geq 1/d$, the most frequent outcome corresponds to a component whose overlap $a_0^2$ with $\ket*{\Psi}$ is comparable to $p_{\max}$, so it can be used as a fiducial state to construct the FSMs $\mathcal{E}_\pm$. Notice that this fiducial state is not necessarily close to $\ket*{\Psi}$, so that $\mathcal{E}_\pm$ are generally non-optimal for the unknown state $\ket*{\Psi}$. However, this guarantees that both non-optimal FSMs together are informationally complete for estimating $\ket*{\Psi}$. Taking into account a sample of size $N_1$, the analytical estimate obtained from Eq. \thinspace \eqref{system_eq} has an infidelity upper bound by $\mathcal{O}(d/a_0^4N_1)$ for almost all states except when $a_0^2\approx1/2$, where it scales in the worst case as $\mathcal{O}(d^3/N_1)$ (SM, Sect. III). Since measurement on an arbitrary basis only ensures $a_0^2\gtrsim 1/d$ (SM, Sec.\ IV), it typically scales as $\mathcal{O}(d^3/N_1)$. These are finite-sample high-probability bounds, valid for any sample size $N_1$, obtained with concentration inequalities for vector-valued random variables \cite{deGois2024} and multinomial random variables \cite{Tomomi2022,Mikosch1996WeakCA} (SM, Sec.\ III). Another observation is that a new POVM with $4d-2$ outcomes can be generated by combining POVMs $\mathcal{E}_\pm$ by flipping a coin with probabilities $p_\pm$. This is a new FSM that estimates states close to the fiducial state and achieves the GMB (SM, Sect.\ II C).
 
The estimate $|\tilde\Psi\rangle$ obtained from the two non-optimal FSMs $\mathcal{E}_\pm$ allows us to perform a near-optimal estimation of $|\Psi\rangle$ by adapting the FSM $\mathcal{E}$. Let $U$ be a unitary operator that transforms the fiducial state $|0\rangle$ into the estimate of the two FSMs, that is, $|\tilde{\Psi}\rangle=U|0\rangle$. The unitary $U$ adapts the FSM $\mathcal{E}$ to $\tilde{\mathcal{E}}=\{U|\varphi^\alpha\rangle\langle\varphi^\alpha|U^\dagger\}_{\alpha=1}^n$, which has as fiducial state $|\tilde\Psi\rangle$, as shown in Fig.\thinspace\ref{Figure1}(c). As long as the estimation with two non-optimal FSMs is performed with a sufficient sample size, the estimate $|\tilde{\Psi}\rangle$ is close enough to the unknown state $|\Psi\rangle$ to ensure that the adapted FSM $\tilde{\mathcal{E}}$ is near-optimal. In fact, the analytical estimate of $\tilde{\mathcal{E}}$ also satisfies a finite-sample bound, whose statistical error scales as $\mathcal{O}(d/N_2)$ for a sample of size $N_2$, while the approximation error described in the previous subsection is now proportional to the infidelity of the first estimate (SM, Sec.\ IV). Then, taking into account this approximation error, the finite-sample bounds guarantee that the adaptation provides an advantage for a large enough sample size $N$. 
 
The sample size $N_0$ used to select the fiducial is negligible relative to $N_1$ and $N_2$, so, for simplicity in the theoretical calculations, we will consider the total sample size to be $N=N_1+N_2$.
 
\subsection{Attaining the Gill-Massar bound with MLE}\label{subsec:bound}
 
The finite-sample bound guarantees an advantage for a large enough $N$, but it does not guarantee that the infidelity achieves the GMB. We show that it can be asymptotically achieved using MLE.
 
For large sample sizes, the MLE is approximately Gaussian distributed around the unknown state \cite{Fujiwara1,Zhu2011,Keenan2025,Almeida2023,Vaart1998,CasellaBerger2002}. Using this approach together with a concentration inequality for Gaussian random variables \cite{Laurent2000}, we derive a high-probability bound on the infidelity of the final estimate $\ket*{\hat\Psi}$ (SM, Sect. V). This bound depends on how the sample size $N$ is divided between $N_1$ and $N_2$. On the one hand, the first stage is less informative than the second stage, since the fiducial state of $\mathcal{E}_\pm$ is far from $\ket*{\Psi}$. On the other hand, a larger sample size in the first stage gives a more accurate first estimate $|\tilde\Psi\rangle$, so that the adapted FSM $\tilde{\mathcal{E}}$ is closer to optimal. Balancing both effects, we obtain the optimal sample size for the first stage,
    \begin{align}
        N_1^*=\left[\frac{(d-1)(2d-1)}{4}\right]^{1/3}N^{2/3}, \label{eq:N1_opt}
    \end{align}
and $N_2^*=N-N_1^*$ for the second stage (SM, Sect. \ V \, D). This optimal splitting is asymptotically valid, that is, for large $N$, and the fraction of the sample size used in the first stage decreases as $N^{-1/3}$. With this splitting, with success probability $1-\delta_1-\delta_2$ we have
\begin{widetext}
    \begin{align}
    \begin{split}
        I(\Psi,\hat\Psi) \lesssim & \frac{d-1}{N}\left\{1+\frac{(d-1)(2d-1)}{2N_1^*}\left[\frac{3}{2}+\sqrt{\frac{2\log(1/\delta_1)}{d-1}}+\frac{\log(1/\delta_1)}{d-1}\right]\right\} \\
        & +\frac{1}{N_2^*}\left(\sqrt{2(d-1)\log\frac{1}{\delta_2}}+\log\frac{1}{\delta_2} \right), \label{eq:bound_MLE}
    \end{split}
    \end{align}
\end{widetext}
where $\delta_1$ and $\delta_2$ are the failure probabilities of the first and second stages, respectively. The first term is the GMB multiplied by a correction due to stage one, which decreases as $1/N_1^*$. This correction multiplies the GMB because the first estimate affects the final estimate only through the Fisher information of $\tilde{\mathcal{E}}$. Substituting $N_1^*$ from Eq.\thinspace\eqref{eq:N1_opt}, we can see that the correction is of the order of $(d^2/N)^{2/3}$. The second term describes the fluctuations of a single estimate around the average, and is of the order of $\sqrt{d}/N$. Therefore, for $N\gg d^2$ the correction vanishes, and the average infidelity approaches the GMB with high probability, that is, the protocol is asymptotically optimal. Another advantage of the MLE is that it does not exhibit the bias of analytical estimators, since it uses the exact probabilities and the combined data of $\mathcal{E}_\pm$ and $\tilde{\mathcal{E}}$ are informationally complete for any pure state with $a_0\neq0$. For smaller sample sizes, the Gaussian description of the MLE may fail, and only the finite-sample bounds of the analytical estimators, described in the previous subsection, are guaranteed.
 
\subsection{Estimation method}
The complete estimation method reads as follows:
	\begin{enumerate}
		\item[(0)] Measurement of a sample of size $N_0\approx 10d$ on an arbitrary basis to select the most frequent outcome as fiducial state, which has a non-negligible overlap $a_0^2\gtrsim 1/d$ with the unknown pure state $|\Psi\rangle$, as shown in Fig.\thinspace\ref{Figure1}(a).
 
		\item[(1)] Acquisition of data through FSMs $\mathcal{E}_\pm$ to obtain probabilities $P_\pm^\alpha$ from a sample size $N_1$. These probabilities are used to obtain a preliminary estimate of $|\Psi\rangle$ by solving Eq.\thinspace\eqref{system_eq}. This estimate serves as a starting point for an MLE \cite{Hradil1997,Shang2017}, which generates the first estimate $|\tilde\Psi\rangle$. This is shown in Fig.\thinspace\ref{Figure1}(b).
 
		\item[(2)] Acquisition of data using an adapted FSM $\tilde{\mathcal{E}}$ to obtain probabilities $\tilde{P}^\alpha$ from a sample size $N_2$. These probabilities, along with those from the previous step $P_\pm^\alpha$, are used to derive a final estimate $|\hat{\Psi}\rangle$ with $N=N_1+N_2$ samples using MLE, with the first estimate $|\tilde{\Psi}\rangle$ serving as the starting point for optimization. This is shown in Fig.\thinspace\ref{Figure1}(c).
	\end{enumerate}

\subsection{Numerical simulations}\label{subsec:sim}
In order to study the performance of the estimation protocol, we carried out several Monte Carlo simulations, the details of which are given in Appendix~\ref{app:mc}. Here we present results for the optimal splitting $N_1=N_1^*$ of Eq.\thinspace\eqref{eq:N1_opt}. Thus, the GMB has the forms $\bar I_{GM_1}=(d-1)/N_1^*$ and $\bar I_{GM_2}=(d-1)/N$ for the first and second stages, respectively.
 
\begin{figure*}[t]\centering
		\includegraphics[width=0.75\linewidth]{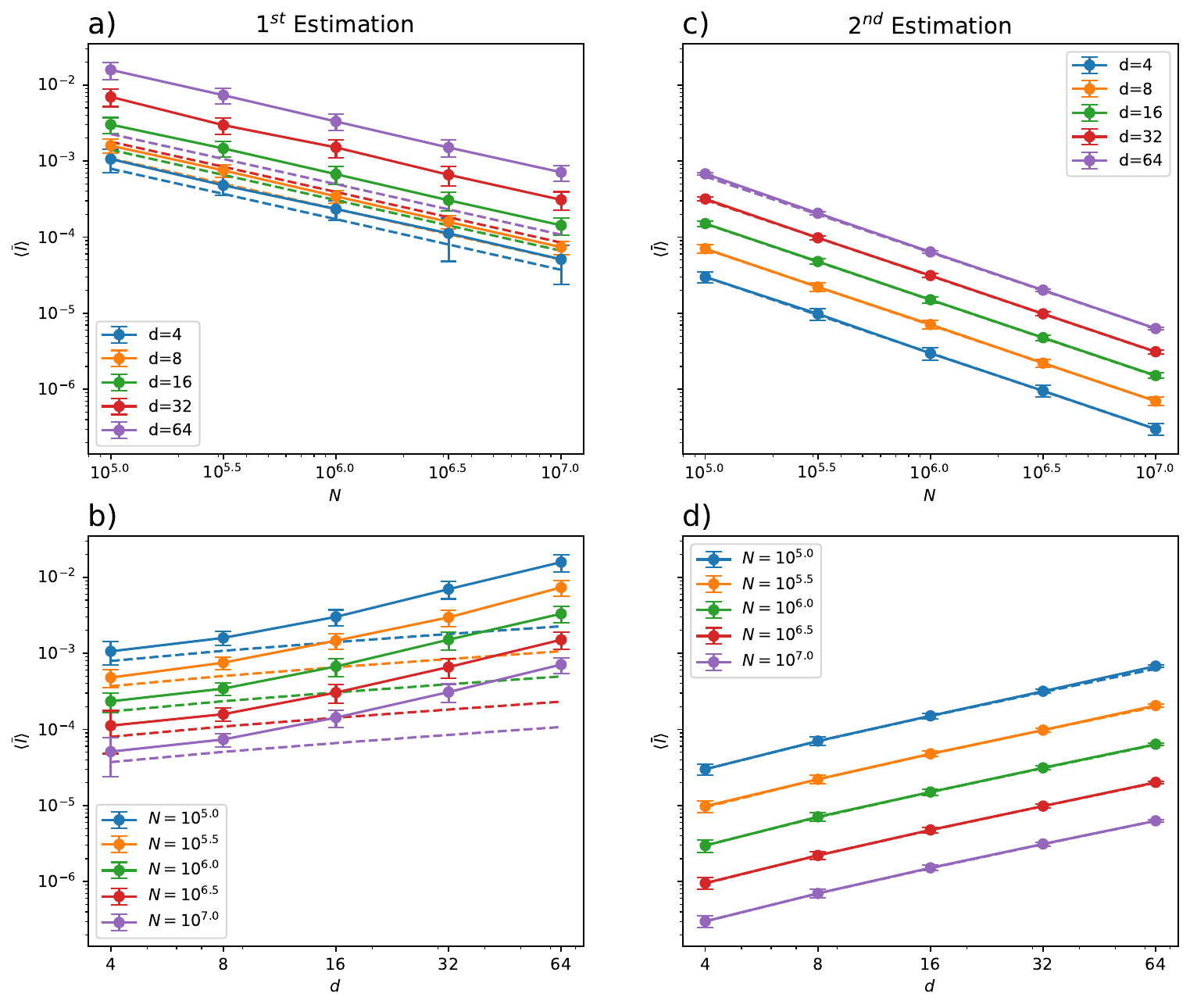}
		\caption{Average infidelity and standard deviation in logarithmic scale for the estimation of pure states, as a function of the sample size $N$ for $N=10^5,10^{5.5},10^6,10^{6.5},10^7$ (insets (a) and (c)) and of the dimension $d=2^n$ for $n=2,3,4,5,6$ qubits (insets (b) and (d)). Insets (a) and (b) correspond to the first stage, and insets (c) and (d) to the second stage. The GMB is displayed (dashed lines) as $\bar I_{GM_1}=(d-1)/N_1^*$ for the first stage and as $\bar I_{GM_2}=(d-1)/N$ for the second stage, with $N_1^*$ the optimal sample size of the first stage. Solid lines are a guide to the eye.}
		\label{Figure2}
	\end{figure*}
 
Figure \ref{Figure2} shows the average of the infidelities $\langle \bar I \rangle $ defined in Appendix~\ref{app:mc} (colored dots) for pure states, as a function of the sample size $N\in\{10^5,10^{5.5},10^6,10^{6.5},10^7\}$ and of the dimension $d\in\{4,8,16,32,64\}$. The error bars correspond to one standard deviation of $\bar I(\Psi_i)$ over $\Omega$. Figures \ref{Figure2}(a) and \ref{Figure2}(b) show the first stage, whose average infidelity is compared to the corresponding GMB $\bar I_{GM_1}$ (colored dashed lines). In Fig.\thinspace\ref{Figure2}(a) we observe that the average infidelity is approximately parallel to the GMB, that is, both decrease as $N^{-2/3}$, since $N_1^*\propto N^{2/3}$. This also suggests the existence of a multiplicative constant. Figure\thinspace \ref{Figure2} (b) shows that the average infidelity increases faster with dimension than with the GMB, indicating a larger exponent of $d$, since the fiducial state of $\mathcal{E}_\pm$ is in general far from the unknown state. The scenario is very different for the second stage, shown in Figs.\thinspace\ref{Figure2}(c) and \ref{Figure2}(d). In both insets, we observe that the average infidelity lies in the GMB $\bar I_{GM_2}$ for all the inspected values of $N$ and $d$, as expected for an asymptotically optimal estimation protocol. Furthermore, the narrow standard deviations indicate that all estimated states reach infidelities close to the GMB. This also holds for $d=64$ and $N=10^5$, although in this case $N/d^2\approx25$, so we are not deep in the asymptotic limit. Thus, the optimal splitting provides a first estimate $|\tilde\Psi\rangle$ that is accurate enough for the adapted FSM $\tilde{\mathcal{E}}$ to be nearly optimal, although the fraction of the sample size used in the first stage decreases as $N^{-1/3}$.
 
In order to obtain the scaling law of the average infidelity of the first and second stages, we fit the function $\bar{I}(\alpha,\beta,\gamma) = \alpha (d-1)^\gamma/N^\beta$ to the data of Fig.\thinspace\ref{Figure2}, with $N$ replaced by $N_1^*$ for the first stage. The fitting procedure and the fitted coefficients are given in Appendix~\ref{app:fit} and Table\thinspace\ref{tab:fit}. For both stages, we obtain $\beta\approx1$ for all dimensions and a coefficient $\gamma$ nearly independent of $N$. Then, averaging the fitted coefficients, the average infidelity is given approximately by
\begin{widetext}
    \begin{equation}
		\langle \bar I_{1} \rangle \approx (0.69\pm0.14) \frac{(d-1)^{1.52\pm0.11}}{(N_{1}^{*})^{1.00\pm0.02}},\qquad\langle \bar I_{2}\rangle\approx (1.07\pm0.07) \frac{(d-1)^{1.01\pm0.03}}{N^{1.00\pm0.02}},\label{eq:scaling_2}
	\end{equation}
\end{widetext}
	for the first and second stages, respectively.
 
Comparing both infidelities, we can see that adaptation improves the dimensional scaling from $d^{1.5}$ to $d$. In addition, both numerical average infidelities outperform the theoretical upper bounds. For the first stage, the exponent $\gamma\approx1.5$ is smaller than the exponent $2$ of the bound $\mathcal{O}(d^2/N_1^*)$, although it increases to $\gamma\approx1.8$ when the dimensions $d=4$ and $8$ are excluded from the fit. For the second stage, the bound predicts the GMB with a correction of the order of $(d^2/N)^{2/3}$, and the coefficients $\alpha$ indicate that this correction is already small for the smallest sample sizes. Therefore, the estimation protocol is close to optimal also outside the asymptotic regime.
 
\begin{figure*}[t]\centering
		\includegraphics[width=0.75\linewidth]{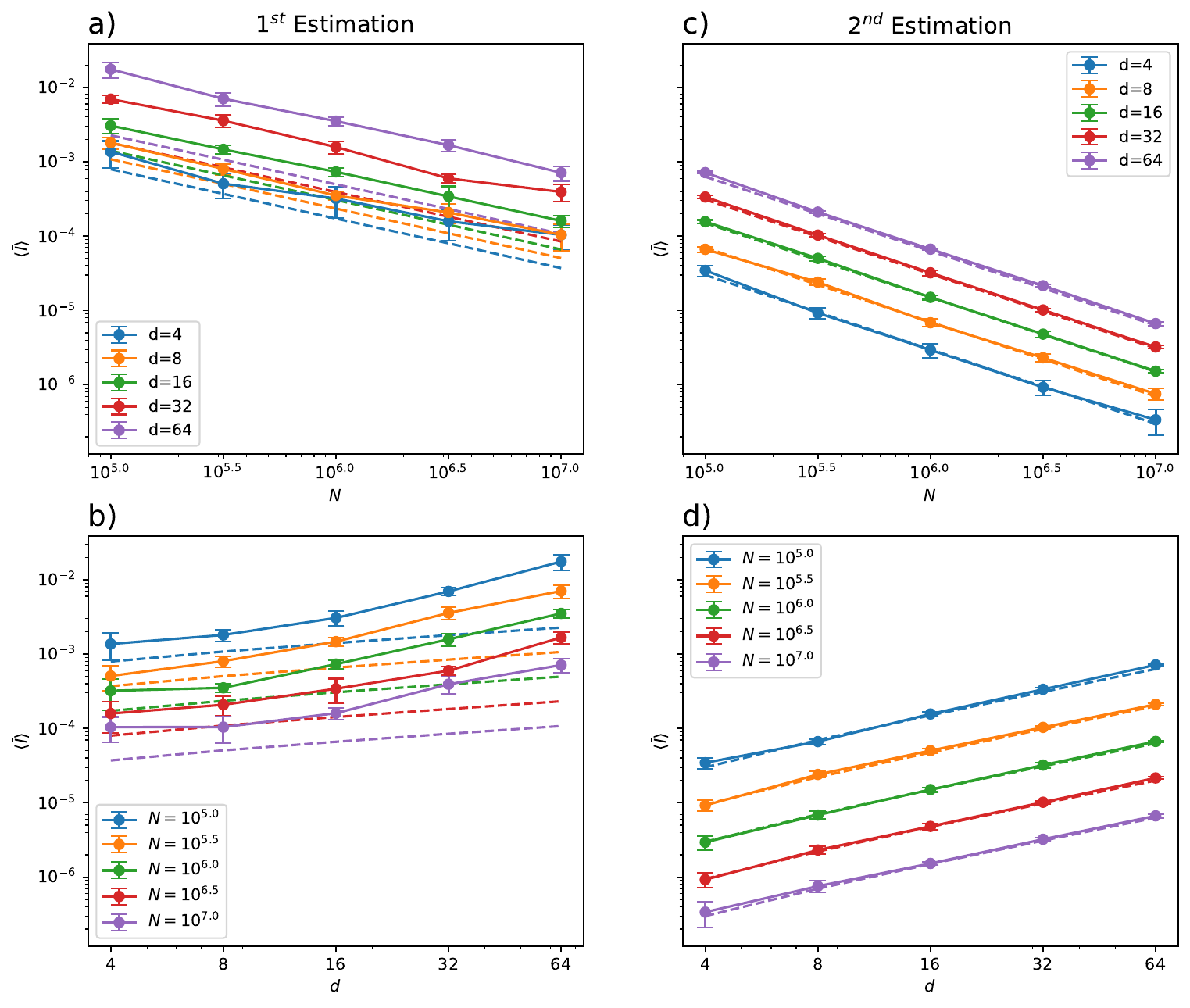}
		\caption{Average infidelity and standard deviation in logarithmic scale for the estimation of weakly mixed states with noise parameter $\lambda=10^{-2}$. The insets and the dashed lines are as in Fig.\thinspace\ref{Figure2}, where the GMB of pure states is displayed only as a reference.}
		\label{Figure3}
	\end{figure*}
 
We also study the estimation of weakly mixed states $\rho=(1-\lambda)\ket*{\Psi}\bra*{\Psi}+\lambda\mathbb{I}/d$, that is, pure states affected by white noise. In this case, a measurement on the computational basis is included to estimate the noise parameter $\lambda$ and to select the fiducial state (SM, Sect.\ VI\,A). With white noise, the Fisher information of the adapted FSM is multiplied by $(1-\lambda)^2$, so that the average infidelity is divided by approximately $(1-\lambda)^2$, while the optimal fraction of Eq. \thinspace \eqref{eq:N1_opt} is multiplied only by $(1-\lambda)^{-2/3}$ (SM, Sect. \ VI \, B). Then, we employ the same sample size $N_1^*$ for mixed states. Figure\thinspace\ref{Figure3} shows the average infidelity $\langle\bar I\rangle$ of both stages for $\lambda=10^{-2}$, in the same way as in Fig. \thinspace \ref{Figure2}. For $\lambda>0$, FSMs do not saturate the Gill-Massar inequality, so the pure-state GMB is only a reference. However, for $\lambda=10^{-2}$ the factor $(1-\lambda)^2$ changes infidelity only by about $2\%$. We observe that the results are very similar to those of pure states. Averaging the fitted coefficients of Table\thinspace\ref{tab:fit_mixed}, for the first stage we obtain $\alpha=0.76\pm0.19$, $\beta=0.94\pm0.09$ and $\gamma=1.44\pm0.13$, and for the second stage $\alpha=1.07\pm0.07$, $\beta=1.00\pm0.02$ and $\gamma=1.01\pm0.03$, that is, the average infidelity is again close to the GMB for all the inspected values of $N$ and $d$, with narrow standard deviations. Therefore, the estimation protocol is robust against weak white noise.

\section{Discussion}\label{sec3}

We introduce a three-stage adaptive estimation protocol for unknown $d$-dimensional pure states, based on FSMs and asymptotically optimal. We derived a high-probability bound on MLE infidelity, used it to determine the optimal splitting of the sample size between stages, and showed that, with this splitting, average infidelity approaches the GMB for $N\gg d^2$. In addition, numerical simulations with this splitting show that the AFSM is close to the GMB across all inspected sample regimes. Thus, our protocol also achieves a nearly optimal estimation accuracy for moderate sample sizes. Previous results for qubits \cite{Gill,Bagan2002,Bagan2006,Hayashi2008} and for pure states of any dimension \cite{Girotti2024} show that the GMB is attained asymptotically, but they do not provide error bounds. Unlike them, our bounds give the optimal splitting of the sample size and the order of the correction to the GMB, that is, they quantify how fast the infidelity approaches the optimal value.
 
Estimation using two non-optimal FSMs and AFSM offers advantages in infidelity scaling and the number of outcomes or bases compared with other well-known estimation methods for pure states \cite{Pereira2022, Zambrano2024, Ma2016, Tariq2024, Wang2022, Carmeli2016, Zambrano12020}. A paradigmatic example is 5-basis tomography (5BBT) \cite{Goyeneche, Zambrano2}, which, with measurements in 5 bases, allows the estimation of a pure state of any dimension with an average infidelity that scales as $\mathcal{O}(d^{1.87}/N)$. The two non-optimal FSMs can be implemented with 4 bases \cite{Li}, one fewer than 5BBT, or with a single FSM having more measurement outcomes. Therefore, the estimate obtained after the first stage has an infidelity that scales approximately as $d^{1.5}$, comparable to the scaling of 5BBT, but with a smaller number of measurement outcomes. Furthermore, the third AFSM can be replaced by two measurement bases, yielding a total of 7 bases. Then, with 2 additional bases beyond the 5 used by the 5BBT, it is possible to reach the GMB asymptotically. Another example is projected least squares tomography \cite{Guta2020}, which employs global 2-designs to estimate rank-one density matrices with an infidelity upper-bounded by $\mathcal{O}(d\log d/N)$, which is $\log d$ times worse than the estimator provided by the AFSM, although the bound of projected least squares holds for any sample size. Moreover, 2-designs require either a $d^2$-outcome SIC-POVM or $d+1$ MUBs, which is much more costly than implementing the AFSMs via 3 POVMs with $2d-1$ outcomes or 7 bases. Recently, it has been shown that non-adaptive Pauli measurements can estimate $n$-qubit pure states with infidelity $\tilde{\mathcal{O}}(d {\rm poly}(n)/N)$ \cite{tomo_pure_binary_tree_2026}. AFSM achieves an asymptotic bound that is better than this proposal by a factor of ${\rm poly}(n)$ and applies to arbitrary dimension $d$, although for $n$ qubits the FSMs require entangling measurements, and the bound for Pauli measurements holds for any sample size. All of this indicates that AFSM requires only a small number of outcomes and bases to yield an asymptotically optimal estimate of unknown pure states, offering a resource-efficient approach for the characterization and benchmarking of quantum hardware. 
	
The GMB can also be reached by a sequence of FSMs, where each FSM uses the previous estimate as its fiducial state, or by a sequence of adapted measurement bases that solve an infidelity-optimization problem \cite{Ferrie, Utreras, Zambrano, granade2017, Farooq2022}. In both cases, many more iterations and measurements are required than in our protocol, which uses just two adaptation steps, one to set the fiducial state and the other to adapt the FSM. It is possible to reduce the number of FSMs even further to one, at the expense of collective measurements across multiple copies of the unknown state, which our protocol does not require. 
 
Integrated photonic processors (IPPs) \cite{Petangelo, Carine2020} are ideal for the experimental testing of AFSMs. These devices are universal and reconfigurable networks of Mach-Zehnder interferometers, fabricated via femtosecond laser writing, which act on spatial qudits encoded in single photons. Universality and reconfigurability of IPPs allow AFSMs to be implemented through measurement bases or POVMs via a direct-sum representation \cite{Martinez2023}. Quantum computers can also implement AFSMs thanks to recent advances in the optimal implementation of POVMs \cite{Yordanov,Pinto,Singal,Ivashkov2024,Stricker2022}.
 
An extension of our method is the estimation of pure states affected by a depolarizing channel. As we show in SM, Sect.\ VI, a measurement on the computational basis allows us to estimate the depolarizing parameter, and then the pure state prior to depolarization is estimated from the statistics of the FSMs. Numerical simulations with the optimal splitting indicate that this state can be estimated close to the GMB in the presence of weak depolarizing noise. Furthermore, it has recently been shown that optimal infidelity scaling can be achieved to estimate a density matrix by first purifying and then applying a tomographic method for rank-one density matrices \cite{2511.15806}. Our method can be combined with this strategy, at the expense of entangling operations across copies, to achieve near-optimal infidelity estimation for arbitrary mixed states with respect to the Gill-Massar bound. Another extension would be the introduction of classical shadows \cite{Aaronson2020,wei2024} to reduce data acquisition complexity, and tensor networks \cite{Cramer2010, Lanyon2017,Baumgratz} or neural networks \cite{Torlai2018,Quek2021} to reduce the computational cost of maximum-likelihood estimation.

\begin{acknowledgments}
C.\thinspace V. and A.\thinspace D. are supported by the ANID Millennium Science Initiative Program ICN17$_-$012.  L.P. was supported by the Government of Spain (Severo Ochoa CEX2019-000910-S, FUNQIP, and QEC4QEA PCI2025-163167), European Union (PASQuanS2.1, 101113690 and QEC4QEA, 101194322), Fundació Cellex, Fundació Mir-Puig, and Generalitat de Catalunya (CERCA program).
\end{acknowledgments}

\section*{Data availability}

The data that support the findings of this study are available in the repository \url{https://github.com/CoVargasRo/AdapFSM}.

\section*{Code availability}

The code used to generate the numerical results of this study is available in the repository \url{https://github.com/CoVargasRo/AdapFSM}.

\appendix
\section{Monte Carlo simulations}\label{app:mc}
 
The Monte Carlo simulations are carried out as follows. We generate a set $\Omega=\{|\Psi_{i}\rangle\}_{i=1}^{100}$ of pure states chosen according to a Haar-uniform distribution for a fixed dimension $d$ \cite{haar}. For each state $|\Psi_{i}\rangle$ we obtain a set $\Lambda_i=\{|\hat\Psi_{i,j}\rangle\}_{j=1}^{10}$ of estimates with our protocol. We consider dimensions $d=2^n$ with $n=2,\dots,6$ and total sample sizes $N\in\{10^5,10^{5.5},10^6,10^{6.5},10^7\}$. The explicit FSM used for the estimation is described in the SM, Sect.\ II\,C. The preliminary stage employs the sample size $N_0=10d$, the first stage employs the sample size $N_1^*$ of Eq.\thinspace\eqref{eq:N1_opt}, divided equally between $\mathcal{E}_+$ and $\mathcal{E}_-$, and the second stage employs the remaining sample size $N-N_1^*-N_0$. The counts of each measurement are simulated using multinomial distributions. For weakly mixed states, we generate a set $\Omega$ of $10$ Haar-random states for each dimension, with noise parameter $\lambda=10^{-2}$, and we obtain $10$ estimates for each state. The computational basis is measured with a sample size $10d$, and the FSMs with the sample sizes given above.

The estimates are used to calculate the average infidelity for each state $|\Psi_i\rangle$ in the set of its estimates as $\bar I(\Psi_i)=\sum_{\Lambda_i}I(\Psi_i,\hat\Psi_{i,j})/|\Lambda_i|$, which is lower bounded by the GMB to leading order in $1/N$. We change $|\hat\Psi_{i,j}\rangle$ for $|\tilde\Psi_{i,j}\rangle$ for the infidelities of the first stage. Similarly, we calculate the average of this infidelity over the set of randomly generated states as $\langle \bar I \rangle = \sum_{\Omega} \bar I(\Psi_i) / |\Omega|$.
 
\section{Maximum likelihood estimation}\label{app:mle}
 
The MLE of the first and second stages maximizes the log-likelihood
\begin{align}
    \log L(\Psi)=\sum_{m}\sum_{\alpha}n_m^\alpha\log p_m^\alpha(\Psi),
\end{align}
where $n_m^\alpha$ is the number of counts of the outcome $\alpha$ of the measurement $m$ and $p_m^\alpha(\Psi)$ its probability. In the first stage, measurements are the computational basis and $\mathcal{E}_\pm$, in the second stage we include $\tilde{\mathcal{E}}$. Thus, the Fisher information of all the data is the sum of the Fisher information of each sample, as required by the analysis of the SM, Sect.\ V \, A. The optimization starts from the analytical estimate of Eq.\thinspace\eqref{system_eq} in the first stage and from the first estimate $|\tilde\Psi\rangle$ in the second stage. For weakly mixed states, the noise parameter $\lambda$ is estimated together with $\ket*{\Psi}$.
 
\section{Fitting procedure}\label{app:fit}
 
The scaling laws of Eq.\thinspace\eqref{eq:scaling_2} are obtained by fitting the function
\begin{align}
    \log_{10}\langle\bar I\rangle=\log_{10}\alpha-\beta\log_{10}N+\gamma\log_{10}(d-1)
\end{align}
to the average infidelities, where $N$ is replaced by $N_1^*$ for the first stage. First, we determine the coefficient $\beta$ for each value of $d$ by fitting a linear function to the curves of $\log_{10}\langle\bar I\rangle$ versus $\log_{10}N$. Next, we set $\beta$ to its average $\bar\beta$ and determine the coefficient $\gamma$ for each value of $N$ by fitting a linear function to the curves of $\log_{10}(N^{\bar\beta}\langle\bar I\rangle)$ versus $\log_{10}(d-1)$. For the first stage, we use $\log_{10}[(N_1^*)^{\bar\beta}\langle\bar I\rangle]$, since $N_1^*$ depends on $d$ through Eq. \thinspace \eqref{eq:N1_opt}. Finally, we set $\gamma$ to its average $\bar\gamma$ and determine $\alpha$ for each value of $d$ using a nonlinear fit of $\alpha(d-1)^{\bar\gamma}/N^{\bar\beta}$ to the curves of $\langle\bar I\rangle$ versus $N$. Tables\thinspace\ref{tab:fit} and \ref{tab:fit_mixed} show the fitted coefficients for the pure and weakly mixed states, respectively. For both kinds of states, we use $\bar\beta=1$ for both stages and $\bar\gamma=1.5$ and $\bar\gamma=1$ for the first and second stages, respectively. The coefficients of Eq.\thinspace\eqref{eq:scaling_2} and those reported for mixed states are the averages of the fitted values, and their uncertainties combine in quadrature the standard deviation of the fitted values and the root mean square of their fitting errors.
 
\begin{table*}[t]
\centering
\caption{Fitted coefficients of the average infidelity $\alpha(d-1)^\gamma/N^\beta$ for pure states. The coefficients $\alpha$ and $\beta$ are fitted for each dimension $d$, and $\gamma$ for each sample size $N$. For the first stage, $N$ is replaced by $N_1^*$.}\label{tab:fit}
\setlength{\tabcolsep}{4pt}
\begin{tabular}{@{}cccccccc@{}}
\toprule
 & & \multicolumn{3}{c}{First stage} & \multicolumn{3}{c}{Second stage}\\
\cmidrule(lr){3-5}\cmidrule(lr){6-8}
$d$ & $\log_{10}N$ & $\alpha$ & $\beta$ & $\gamma$ & $\alpha$ & $\beta$ & $\gamma$\\
\midrule
4 & 5 & $0.770\pm0.005$ & $0.98\pm0.01$ & $1.6\pm0.1$ & $1.13\pm0.03$ & $1.00\pm0.03$ & $1.01\pm0.04$\\
8 & 5.5 & $0.559\pm0.001$ & $1.007\pm0.004$ & $1.5\pm0.1$ & $0.96\pm0.02$ & $0.98\pm0.02$ & $1.02\pm0.02$\\
16 & 6 & $0.558\pm0.004$ & $1.00\pm0.01$ & $1.5\pm0.1$ & $1.043\pm0.003$ & $1.008\pm0.006$ & $1.03\pm0.01$\\
32 & 6.5 & $0.69\pm0.01$ & $1.01\pm0.02$ & $1.5\pm0.1$ & $1.080\pm0.005$ & $1.008\pm0.003$ & $1.03\pm0.01$\\
64 & 7 & $0.873\pm0.005$ & $1.02\pm0.01$ & $1.5\pm0.1$ & $1.13\pm0.01$ & $1.01\pm0.01$ & $0.98\pm0.01$\\
\bottomrule
\end{tabular}
\end{table*}
 
\begin{table*}[t]
\centering
\caption{Fitted coefficients of the average infidelity $\alpha(d-1)^\gamma/N^\beta$ for weakly mixed states with $\lambda=10^{-2}$, as in Table\thinspace\ref{tab:fit}.}\label{tab:fit_mixed}
\setlength{\tabcolsep}{4pt}
\begin{tabular}{@{}cccccccc@{}}
\toprule
 & & \multicolumn{3}{c}{First stage} & \multicolumn{3}{c}{Second stage}\\
\cmidrule(lr){3-5}\cmidrule(lr){6-8}
$d$ & $\log_{10}N$ & $\alpha$ & $\beta$ & $\gamma$ & $\alpha$ & $\beta$ & $\gamma$\\
\midrule
4 & 5 & $0.96\pm0.04$ & $0.82\pm0.07$ & $1.5\pm0.1$ & $1.13\pm0.03$ & $1.00\pm0.03$ & $1.01\pm0.04$\\
8 & 5.5 & $0.62\pm0.01$ & $0.92\pm0.04$ & $1.5\pm0.1$ & $0.96\pm0.02$ & $0.98\pm0.02$ & $1.02\pm0.02$\\
16 & 6 & $0.57\pm0.01$ & $0.96\pm0.01$ & $1.5\pm0.1$ & $1.043\pm0.003$ & $1.008\pm0.005$ & $1.030\pm0.006$\\
32 & 6.5 & $0.71\pm0.01$ & $0.99\pm0.06$ & $1.4\pm0.1$ & $1.080\pm0.005$ & $1.008\pm0.003$ & $1.03\pm0.01$\\
64 & 7 & $0.95\pm0.03$ & $1.03\pm0.03$ & $1.3\pm0.1$ & $1.13\pm0.01$ & $1.010\pm0.008$ & $0.98\pm0.01$\\
\bottomrule
\end{tabular}
\end{table*}

\bibliography{bibtex}

%%%%%%%%%%%%%%%%%%%%%%%%%%%%%%%%%%%%%%%%%%%%%%%%%%%%%%%%%%%%%%%%%%%%%%%
%%%%%%%%%%%%%%%%%%%%%%%% SUPPLEMENTARY INFORMATION %%%%%%%%%%%%%%%%%%%%
%%%%%%%%%%%%%%%%%%%%%%%%%%%%%%%%%%%%%%%%%%%%%%%%%%%%%%%%%%%%%%%%%%%%%%%
\clearpage
\onecolumngrid
\begin{center}
\textbf{\large Supplementary Material}
\end{center}

% leave appendix mode: restore the regular section style and numbering
\makeatletter
\let\@hangfrom@section\SM@hangfrom@section
\let\@sectioncntformat\@seccntformat
\@removefromreset{equation}{section}
\def\theequation@prefix{}
\makeatother
\renewcommand{\thesection}{\Roman{section}}
\renewcommand{\thesubsection}{\Alph{subsection}}
\renewcommand{\thesubsubsection}{\arabic{subsubsection}}
\setcounter{section}{0}
\setcounter{equation}{0}
\setcounter{figure}{0}
\setcounter{table}{0}
\renewcommand{\theequation}{S\arabic{equation}}
\renewcommand{\thefigure}{S\arabic{figure}}
\renewcommand{\thetable}{S\arabic{table}}
\renewcommand{\theHequation}{S\arabic{equation}}
\renewcommand{\theHfigure}{S\arabic{figure}}
\renewcommand{\theHtable}{S\arabic{table}}
\renewcommand{\theHsection}{S\arabic{section}}

\let\addcontentsline\SIaddcontentsline
\tableofcontents

%-------------------------------------------------------------------------------
\section{Quantum State Estimation}

This section presents an introduction to the estimation of the quantum state, which refers to a collection of statistical methods used to reconstruct the density matrix of a quantum system from measurement data \cite{Teo2015}. Furthermore, it establishes fundamental limits on the accuracy of such estimations through the Cram\'er-Rao bound and the Gill-Massar inequality.

\subsection{Cram\'er-Rao bound}

Consider a $d$-dimensional quantum system in a state $\rho(\bm{s})$ characterized by the parameters $\bm{s} = \mqty[s_1, \cdots, s_n]^{\top}$, which we aim to estimate. Given a set of measurement operators $\{E_i\}$ that satisfies the completeness relation $\sum_i E_i = I$, the probability of obtaining the outcome $i$ in a measurement is given by
\begin{align}
    p(i|\bm{s})= \Tr \big(E_i\rho(\bm{s})\big).
    \label{Qprob}
\end{align} 
From the experiment, we obtain estimates $\bm{\hat{s}}(x)= [\hat{s}_1(x), \cdots, \hat{s}_n(x)]^{\top}$ of the original parameters, which are functions of the experimental results. The uncertainty of the estimation of the parameters $\bm{s}$ is given by the covariance matrix,
\begin{align}
    \mathbb{C}_{ij}= \Cov(\hat{s}_i,\hat{s}_j)=\mathbb{E}\left[ \left( \hat{s}_i -\mathbb{E}(\hat s_i)  \right)\left(\hat{s}_j -\mathbb{E}(\hat s_j)  \right)  \right].
\end{align}
The Cram\'er-Rao inequality imposes a fundamental lower bound for $\mathbb{C}$. This is given by
\begin{align}
    \mathbb{C}\geq G^{\top}C^{-1}G,
\end{align}
where $G_{ij} = \mathbb{E}\left(\hat{s}_j(x)\pdv{}{s_i}\ln p(x|\bm{s}) \right)$ and $C$ is the classical Fisher information matrix
\begin{align}
    C_{ij}=\mathbb{E}\left(  \pdv{}{s_i}\ln p(x|\bm{s})\pdv{}{s_j}\ln p(x|\bm{s})   \right).
    \label{fishermatrix}
\end{align}
This can be used to measure how much information about $\bm{s}$ is likely to be obtained from a random experiment. A specific case of this inequality arises when the estimator is unbiased, that is, $\mathbb{E}[\bm{\hat s}(x)]=\bm{s}$. Then, the Cram\'er-Rao bound for an unbiased estimator becomes
\begin{equation}
    \mathbb{C}\geq C^{-1}.
\end{equation}
We will now extend this result to the quantum case. Substituting the probability distribution \eqref{Qprob} into the classical Fisher information matrix (\ref{fishermatrix}), we obtain
\begin{align}
    C_{ij} &= \sum_{k=1}^{m} \frac{1}{p(k|\bm{s})} \Tr\Big(E_k\pdv{\rho(\bm{s})}{s_i}  \Big) \Tr\Big(E_k\pdv{\rho(\bm{s})}{s_j}  \Big). 
    \label{clasicFisher}
\end{align}
From this equation, we can see that the classical Fisher information matrix $C$ depends on the quantum state $\rho$, but also the set of measurements $\{E_k\}$. Maximizing $C$ on the set of quantum measurements $\{E_k\}$ we obtain an upper bound for the classical Fisher information matrix $C$,
\begin{align}
C \leq Q, \label{desigualdad_I_J}
\end{align}
where $Q$ is the quantum Fisher information matrix. This is explicitly given by
\begin{align}
Q_{ij}(\bm{s})=\frac{1}{2}\Tr\Bigg(\rho(\bm{s})\Big[L_i(\bm{s}) L_j(\bm{s})     +L_j(\bm{s}) L_i(\bm{s}) \Big] \Bigg), \label{QFI}
\end{align}
where  $L_i (\bm{s})$ are the \textit{Symmetric Logarithmic Derivative} (SLD) operators, which are implicitly defined by 
\begin{align}
    \pdv{\rho(\bm{s})}{s_i} = \frac{1}{2} \Big(\rho(\bm{s}) L_i(\bm{s}) + L_i(\bm{s}) \rho(\bm{s})\Big).
    \label{SLD}
\end{align}  
The quantum Fisher information matrix $Q$ depends only on the quantum state of the system. It is the maximum amount of information that can be retrieved from the system via quantum measurements.

The Quantum Cram\'er-Rao bound establishes a fundamental limit on the precision of parameter estimation in quantum systems. This bound is expressed as an inequality for the covariance matrix $\mathbb{C}$ of an unbiased estimator:  
\begin{align}
    \mathbb{C} \geq C^{-1} \geq Q^{-1}.
    \label{QCR}
\end{align}

Let $\mathbb{C}^{(N)}(\bm{\hat s})$ be the covariance matrix and $C^{(N)}(\bm{s})$ be the classical Fisher information matrix for an estimate with sample size $N$. The quantum Cram\'er-Rao bound becomes
\begin{align}
    \mathbb{C}^{(N)}\geq\big[ C^{(N)}\big]^{-1}\geq\frac{1}{N}Q^{-1},
    \label{cramer-rao-N}
\end{align}
where $Q$ corresponds to the quantum Fisher information matrix for a single copy.

\subsection{Gill-Massar inequality}

    The quantum Cram\'er-Rao bound sets the fundamental lower limit over all possible quantum measurements. However, it has been shown that in the context of quantum state estimation using individual measurements, that is, performing separate measurements on each copy of the state $\rho$, this bound cannot generally be saturated. In particular, no single POVM can produce a classical Fisher information matrix that matches the quantum Fisher information matrix for all parameters \cite{Gill}. 

    In fact, the Gill-Massar inequality establishes a fundamental limitation on state estimation. For pure states it holds for arbitrary measurements, including collective measurements on many copies \cite{Gill}, while for mixed states it holds for separable (individual) measurements. It states that
    \begin{equation}
        \Tr\left(C^{(N)} Q^{-1}\right) \leq N(d-1),
        \label{gillmassar}
    \end{equation}
    where $d$ is the dimension of the Hilbert space to which $\rho$ belongs. Using this inequality, we can also derive a bound for the weighted mean squared error $\Tr(W\mathbb{C})$, with $W$ being a positive weight matrix and $\mathbb{C}$ the covariance matrix \cite{ZhuThesis}. This is given by
    \begin{align}
        \Tr(W \mathbb{C}) &\geq \frac{1}{N(d-1)} \left(\Tr\sqrt{Q^{-1/2}W Q^{-1/2}}\right)^2 .
        \label{GM for MSE}
    \end{align}
    The minimum is achieved when $\mathbb{C}=(C^{(N)})^{-1}$ and when
    \begin{align}
         C^{(N)} = N(d-1)Q^{1/2}\frac{\sqrt{Q^{-1/2}WQ^{-1/2}}}{\Tr\sqrt{Q^{-1/2}WQ^{-1/2}}}Q^{1/2}.
    \end{align}
    Taking into account $W = Q(\rho)/4$, the weighted mean squared error becomes the mean Bures distance, which is related to the mean infidelity $\bar{\mathcal{I}}$ as
    \begin{align}
        \Tr(W\mathbb{C}) = \bar{\mathcal{I}}(\rho_1, \rho_2) &\geq \frac{n^2}{4N(d-1)},\qquad C= \frac{(d-1)}{n} Q,
    \end{align}
    with $n$ the number of parameters that are estimated, $d$ the dimension of the Hilbert space of the states to be estimated, and $N$ the sample size. If we restrict ourselves to the estimation of pure states, which are defined by $n=2(d-1)$ real parameters, the lower bound of the mean infidelity becomes
    \begin{equation}
        \bar{\mathcal{I}}(|\psi\rangle)_{pure} \geq \frac{d-1}{N}.
    \end{equation}
    Thereby, we can use the average infidelity as a metric for the error in estimating pure states.
    
%---------------------------------------------------------------------------------
    
\section{Fisher Symmetric Measurements}\label{apendix1}
	
	Consider an unknown quantum pure state $\ket*{\Psi(\bm{x})}$ that depends on a vector of real parameters $\bm{x}$ and is close to a known fiducial pure state $\ket*{0}$. This state can be written as
	\begin{equation}
		\ket*{\Psi(\bm{x})} = a_0(\bm{x})\ket*{0} + \sum_{j=1}^{d-1}(x_{j0} + ix_{j1})\ket*{j}, \label{eq:state}
	\end{equation}
	with $\bm{x}=[x_{0,0},x_{0,1},\cdots, x_{d-1,0}, x_{d-1,1}]^\top$ and $|x_{k\sigma}| \ll 1$ infinitesimal real parameters. The parameter $a_0\approx 1$ is the overlap with the fiducial state and is explicitly given by
    \begin{align}
        a_0(\bm{x}) = \sqrt{1-\sum_{j=1}^{d-1}\sum_{\sigma=0}^1 (x_{j\sigma})^2}.
    \end{align}
    Writing the pure state $\ket*{\Psi(\bm{x})}$ as a density matrix and retaining only the linear order in $x_{j\sigma}$, we obtain
	\begin{equation}
		\rho(\bm{x}) = \ket*{0}\bra*{0} + \sum_{j\sigma}x_{j\sigma}X_{j\sigma}, \label{eq:rho_approx}
	\end{equation}
	where
	\begin{equation}
		X_{j\sigma} = (-i)^\sigma (\ket*{0}\bra*{j} + (-1)^{\sigma}\ket*{j}\bra*{0}).
	\end{equation}
	These hermitian operators satisfy the orthogonality condition $\Tr(X_{j\sigma}X_{k\tau})=2\delta_{jk}\delta_{\sigma\tau}$. For simplicity, we are going to omit the dependence on $\bm{x}$ except when it is strictly necessary. \\
	
	A positive operator-valued measure (POVM) \cite{Nielsen} is a Fisher symmetric measurement (FSM) \cite{Li} if it saturates the Gill-Massar Inequality \cite{Gill}
	\begin{equation}
		\Tr(Q^{-1}C) = d-1,
	\end{equation}
	and the corresponding classical Fisher information matrix (FIM) is as evenly distributed as possible over all parameters. Both conditions are fulfilled when the classical FIM $C$ is related to the quantum FIM $Q$ by \cite{Li}
	\begin{align}
		C = \frac{1}{2}Q.
	\end{align}
	In the following sections, we calculate both quantum and classical Fisher information matrices for the state \eqref{eq:state} and find the conditions for a POVM to be a FSM.
	
	\subsection{Quantum Fisher Information}\label{sec:qfi}
	
	The quantum FIM of the state $\rho$ is a symmetric matrix $Q$ with elements
	\begin{equation}
		Q_{j\sigma,k\tau} = \frac{1}{2}\Tr[\rho(L_{j\sigma} L_{k\tau} + L_{k\tau} L_{j\sigma} )],
	\end{equation}
	where $\{L_{j\sigma}\}$ are the symmetric logarithmic derivative operators, implicitly determined by the equation 
	\begin{align}
		\frac{\partial \rho(\bm{x})}{ \partial x_{j\sigma} } = \frac{1}{2}(L_{j\sigma}\rho + \rho L_{j\sigma} ).
	\end{align}
	When the state is pure, that is, $\rho(\bm{x})=|\psi\rangle\langle\psi|$, the quantum Fisher information matrix is simply given by
	\begin{align}
		Q_{j\sigma,k\tau}= 4 {\rm Re} \left[\langle \partial_{j\sigma} \psi | \partial_{k\tau} \psi\rangle - \langle \psi | \partial_{j\sigma} \psi\rangle \langle \partial_{k\tau} \psi |\psi \rangle  \right],
	\end{align}
	where 
	\begin{align}
		|\partial_{j\sigma} \psi\rangle = \frac{\partial}{\partial x_{j\sigma} } |\psi\rangle.
	\end{align}
	Thus, the quantum Fisher information matrix of the state $|\Psi(x)\rangle$ is
	\begin{align}
		Q_{j\sigma,k\tau} = 2\Tr(X_{j\sigma} X_{k\tau}) = 4\delta_{j\sigma,k\tau},
	\end{align}
	or equivalently $Q(\rho) = 4\mathbb{I}_{2d-2}$. 
	
	\subsection{ Classical Fisher Information}
	
	Consider an arbitrary POVM with $n$ rank-1 elements, $\mathcal{E}=\{ E^\alpha=\ket*{\varphi^\alpha}\bra*{\varphi^\alpha} \}$, where
	\begin{equation}
		\ket*{\varphi^\alpha} = \sum_{k=0}^{d-1}(\beta_k^\alpha + i\gamma_k^\alpha)\ket*{k}.
	\end{equation}
	with $\{\beta_k^\alpha\}$ and $\{\gamma_k^\alpha\}$ being real coefficients. Taking into account the completeness condition, $\sum_{\alpha}E^{\alpha} = \mathbb{I}$, we have
	\begin{align}
		\sum_{\alpha}E^{\alpha} &= \sum_{\alpha}\sum_{k,j}(\beta_k^\alpha + i\gamma_k^\alpha)(\beta_j^\alpha - i\gamma_j^\alpha)\ket*{k}\bra*{j}\\
		&= \sum_{\alpha}\sum_{k,j}\left(\beta_k^\alpha \beta_j^\alpha + \gamma_k^\alpha \gamma_j^{\alpha} +i(\beta_j^\alpha\gamma_k^\alpha-\beta_k^\alpha \gamma_j^{\alpha})\right)\ket*{k}\bra*{j} \equiv \sum_{k,j} \delta_{k,j}\ket*{k}\bra*{j}.
	\end{align}
	Defining the $n$-dimensional real vectors $\bm{\beta}_k = [\beta_k^1, \dots, \beta_k^n]^\top$ and $\bm\gamma_k = [\gamma_k^1, \dots, \gamma_k^n]^\top$, the completeness condition becomes
	\begin{align}
		\begin{split}
			\bm\beta_k\cdot \bm\beta_j + \bm\gamma_k\cdot \bm\gamma_j &= \delta_{k,j} ,\quad j,k = 0,\dots,d-1,\\
			\bm\gamma_k \cdot \bm\beta_j -\bm\beta_k\cdot \bm\gamma_j &=  0 ,\quad j,k = 0,\dots,d-1.
		\end{split}\label{eq:completitud}
	\end{align}
    where the centered dot represents the inner product of the real vectors. Since we can choose the global phase of each POVM vector $\ket*{\varphi^\alpha}$, we can set the constant accompanying the vector $\ket*{0}$ to be real, i.e. $\gamma_0=0$,
	\begin{equation}
		\ket*{\varphi^\alpha} = \beta_0^\alpha\ket*{0} + \sum_{k=1}^{d-1}(\beta_k^\alpha + i\gamma_k^\alpha)\ket*{k}. \label{eq:proj_FSM}
	\end{equation}
	With this, we also have the following conditions for $\beta_0$, where we define the real vector $\bm\beta_0=[\beta_0^1,\cdots,\beta_0^n]^\top$
	\begin{equation}
		\begin{split}
			\bm\beta_0\cdot \bm\beta_0 &= 1\\
			\bm\beta_0 \cdot \bm\beta_j &= \bm\beta_0 \cdot \bm\gamma_j = 0 ,\quad j=1,...,d-1.   
		\end{split}
	\end{equation}
	
	The classical FIM of the state $\ket*{\Psi}$ when measured by $\mathcal{E}$ is
    \begin{equation}
		C_{j\sigma, k\tau} = \sum_\alpha \frac{1}{p(\alpha|x)}\frac{\partial p(\alpha|x)}{\partial x_{j\sigma}}\frac{\partial p(\alpha|x)}{\partial x_{k\tau}}, \label{Eq:CQFIM}
	\end{equation}
	where $p(\alpha|x)$ is the probability distribution. Calculating the probabilities for the POVM $\mathcal{E}$, we obtain
	\begin{align}
		p(\alpha|x) &= \Tr(E^\alpha \rho(x)), \nonumber \\
		&= \Tr\left[ E^\alpha \left(\ket*{0}\bra*{0} + \sum_{j\sigma}x_{j\sigma}X_{j\sigma}\right)\right],  \\
		&= \bra*{0}E^\alpha\ket*{0} + \sum_{j\sigma} x_{j\sigma}(-i)^\sigma[\bra*{j}E^\alpha\ket*{0} + (-1)^\sigma \bra*{0}E^\alpha\ket*{j}].
	\end{align}
	Calculating the matrix elements of each operator $E^\alpha$,
	\begin{align}
		\bra*{0}E^\alpha\ket*{0} &= (\beta_0^\alpha + i\gamma_0^\alpha)(\beta_0^\alpha - i\gamma_0^\alpha) = (\beta_0^\alpha)^2,\\
		\bra*{j}E^\alpha\ket*{0} &= (\beta_j^\alpha + i\gamma_j^\alpha)(\beta_0^\alpha - i\gamma_0^\alpha) = \beta_0^\alpha(\beta_j^\alpha + i\gamma_j^\alpha), \\
		\bra*{0}E^\alpha\ket*{j} &= (\beta_0^\alpha + i\gamma_0^\alpha)(\beta_j^\alpha - i\gamma_j^\alpha) = \beta_0^\alpha(\beta_j^\alpha - i\gamma_j^\alpha) . 
	\end{align}
	we can rewrite the probabilities as
	\begin{align}
		p(\alpha|x) &= (\beta_0^\alpha)^2 + 2\beta_0^\alpha\sum_j\left(  x_{j0}\beta_j^\alpha + x_{j1}\gamma_j^\alpha  \right) \label{eq:prob_fsm_opt}
	\end{align}
	Thereby, the derivatives of the probabilities are given by
	\begin{align}
		\frac{\partial p(\alpha|x)}{\partial x_{j0}} = 2\beta_0^\alpha\beta_j^\alpha\\
		\frac{\partial p(\alpha|x)}{\partial x_{j1}} = 2\beta_0^\alpha\gamma_j^\alpha\\
	\end{align}
	Replacing $	p(\alpha|x)$ and $\partial p(\alpha|x)/\partial x_{j\sigma}$ in Eq.\eqref{Eq:CQFIM} and approximating to first order, we obtain the classical FIM
	\begin{equation} 
		\begin{split}
			C_{j0,k0} &= 4\bm\beta_j\cdot \bm\beta_k,  \\
			C_{j1,k1} &= 4\bm\gamma_j\cdot \bm\gamma_k, \\
			C_{j0,k1} &= 4\bm\beta_j\cdot \bm\gamma_k ,    
		\end{split}
		\quad j,k=1,...,d-1.
	\end{equation}

	\subsection{ Explicit construction of FSM}\label{sec:exp_cons_fsm}
    Imposing the relation between classical and quantum FIMs $C=Q/2$, we obtain the conditions
	\begin{align}
		2\bm\beta_j\cdot \bm\beta_k &= \delta_{jk},\\
		2\bm\gamma_j\cdot \bm\gamma_k &= \delta_{jk}, \\
		2\bm\beta_j\cdot \bm\gamma_k &= 0 ,
	\end{align}
    for $j,k = 1,\dots,d-1$. Equivalently,
	\begin{equation}
		\begin{split}
			\bm\beta_0 \cdot \bm\beta_0 &= 1, \\
			\bm\beta_0 \cdot \bm\beta_j &= \bm\beta_0 \cdot \bm\gamma_j = \bm\beta_j \cdot \bm\gamma_k = 0,\\
			\bm\beta_j \cdot \bm\beta_k &= \bm\gamma_j \cdot \bm\gamma_k = \frac{1}{2}\delta_{jk}.  
		\end{split}
		\label{FSMconditions}
	\end{equation}
	Thus, the set of $2d-1$ real vectors $\{\bm\beta_0, \sqrt{2}\bm\beta_k, \sqrt{2}\bm\gamma_k\}_{k=1}^{d-1}$ forms an orthonormal set. Consequently, the number of elements of the POVM must satisfy $n\geq 2d-1$ since the dimension of the vectors must be equal to or greater than the number of states to have an orthonormal set. This defines an efficient procedure for the construction of FSMs \cite{Li}, which involves generating a set of $n$ real vectors of dimension $2d-1$ and using their components to construct the vectors $\ket*{\varphi^\alpha}$. Following this procedure, we can construct the following minimal FSM with fiducial state $\ket*{0}$,
	\begin{align}
		\ket*{\varphi^0} &= \frac{1}{\sqrt{n}}\left[\ket*{0} \pm e^{i\pi/4}\sum_{j=1}^{d-1}\ket*{j} \right],\\
		\ket*{\varphi^{2k-1}} &= \frac{1}{\sqrt{n}}\left[\ket*{0} \pm e^{i\pi/4}\left(z\ket*{k} + \frac{1}{\sqrt{n}+1}\sum_{k\neq j=1}^{d-1}\ket*{j} \right)\right], \\
		\ket*{\varphi^{2k}} &= \frac{1}{\sqrt{n}}\left[\ket*{0} \pm e^{i\pi/4}\left(z^*\ket*{k} + \frac{1}{\sqrt{n}+1}\sum_{k\neq j=1}^{d-1}\ket*{j} \right)\right], 
	\end{align}
	where $k=1,...,d-1$ and $z=\frac{1}{\sqrt{n}+1} - \sqrt{\frac{n}{2}}e^{-i\pi/4}$.

    In the following subsection, we present some side results about FSMs: 1) orthogonality relation between the coefficients of FSM, 2) including an arbitrary local phase in an FSM also yields an FSM, and 3) the combination of several FSMs is also a FSM. 

    \subsubsection{Orthogonality relation}\label{sec:ortho_cond}
    
    The conditions in Eq.\thinspace\eqref{eq:completitud} and \eqref{FSMconditions} can be written more compactly, considering the complex coefficients $\omega_k^\alpha = \beta_k^\alpha +i\gamma_k^\alpha$ as
	\begin{align}
		\sum_{\alpha=1}^{n} \omega_k^\alpha \omega_j^\alpha =& 0,  \quad j,k=1,\dots,d-1 , \label{eq:cond1}\\
		\sum_{\alpha=1}^{n} \omega_k^\alpha \omega_j^{\alpha*} =& \delta_{jk}, \quad j,k=0,\dots,d-1.   \label{eq:cond2}
	\end{align}
    The first equation represents the orthogonality condition between the real vectors used to construct the FSM, while the second is the completeness relation. Moreover, for $n>2d-1$ we can complete the basis $\{ \bm\beta_0,\sqrt{2}\bm\beta_j, \sqrt{2}\bm\gamma_j  \}$ with $n-2d+1$ real vectors $\bm\nu_j$. Then, this new basis satisfies the following orthogonality relation,
\begin{align}
	\beta_0^\alpha \beta_0^{\alpha'} + 2 \sum_{j=1}^{d-1} ( \beta_j^\alpha\beta_j^{\alpha'} + \gamma_j^\alpha\gamma_j^{\alpha'}  ) + \sum_{j=1}^{n-2d+1} \nu_{j}^\alpha \nu_j^{\alpha'} = \delta_{\alpha,\alpha'}.
\end{align}
This expression leads to an upper bound for the norm of the elements of a FSM,
\begin{align}
	| \ket*{\varphi^\alpha} |_2^2 = (\beta_0^\alpha)^2 + \sum_{k=1}^{d-1}[ (\beta_j^\alpha)^2 + (\gamma_j^\alpha)^2 ] \leq \frac{1+(\beta_0^\alpha)^2 }{2}. \label{eq:bound_norm_fsm_elem}
\end{align}
    
    \subsubsection{Arbitrary Phase FSM}\label{apendix2}
	
	In this section, we demonstrate that if $\mathcal{E}=\{ \ket*{\varphi^\alpha}\bra*{\varphi^\alpha} \}$ is an FSM, with $\ket*{\varphi^\alpha}$ given by Eq.\thinspace\eqref{eq:proj_FSM}, then POVM $\mathcal{E}'=\{ \ket*{\varphi'^\alpha}\bra*{\varphi'^\alpha} \}$, with
	\begin{align}
		\ket*{\varphi'^\alpha} &= \beta_0^\alpha \ket*{0} + e^{i \phi} \sum_{k=1}^{d-1} (\beta_k^\alpha + i \gamma_k^\alpha) \ket*{k}, \quad \phi\in \mathbb{R},
	\end{align}
	is also an FSM. Let us define the new coefficients
	\begin{align}
		\beta_k'^\alpha &= \text{Re}\left( e^{i \phi} (\beta_k^\alpha + i \gamma_k^\alpha) \right) = \cos(\phi) \beta_k^\alpha - \sin(\phi) \gamma_k^\alpha. \\
		\gamma_k'^\alpha &= \text{Im}\left( e^{i \phi} (\beta_k^\alpha + i \gamma_k^\alpha) \right) = \cos(\phi) \gamma_k^\alpha + \sin(\phi) \beta_k^\alpha,
	\end{align}
	and their respective real vectors $\bm\beta'_k = (\beta'^0_k, \dots, \beta'^n_k )$ and $\bm\gamma'_k = (\gamma'^0_k, \dots, \gamma'^n_k )$.
	Vectors $\bm\beta_k'$ and $\bm\gamma_k'$ satisfy the FSM conditions. In fact,
	\begin{align}
		\bm\beta_0 \cdot \bm\beta_0 &= 1, \\
		\bm\beta_0 \cdot \bm \beta_k' &= (\cos(\phi) \bm\beta_0 \cdot \bm\beta_k - \sin(\phi) \bm\beta_0 \cdot \bm\gamma_k) = 0, \\
		\bm\beta_0 \cdot \bm\gamma_k' &= (\sin(\phi) \bm\beta_0 \cdot \bm\beta_k + \cos(\phi) \bm\beta_0 \cdot \bm\gamma_k) = 0,\\
		\bm\beta_j' \cdot \bm\gamma_k' &= (\cos(\phi) \bm\beta_j - \sin(\phi) \bm\gamma_j) \cdot (\sin(\phi) \bm\beta_k + \cos(\phi) \bm\gamma_k), \nonumber\\
		&= (\cos(\phi) \sin(\phi) \bm\beta_j\cdot \bm\beta_k + \cos^2(\phi) \bm\beta_j\cdot \bm\gamma_k - \sin^2(\phi) \bm\gamma_j\cdot \bm\beta_k - \cos(\phi) \sin(\phi) \bm\gamma_j\cdot \bm\gamma_k) = 0,\\
		\bm\beta_j' \cdot \bm\beta_k' &= (\cos(\phi) \bm\beta_j - \sin(\phi) \bm\gamma_j) \cdot (\cos(\phi) \bm\beta_k - \sin(\phi) \bm\gamma_k), \nonumber\\
		&= (\cos^2(\phi) \bm\beta_j \cdot \bm\beta_k - \sin(\phi) \cos(\phi) \bm\gamma_j\cdot \bm\beta_k - \sin(\phi) \cos(\phi) \bm\beta_j\cdot \bm\gamma_k + \sin^2(\phi) \bm\gamma_j\cdot \bm\gamma_k) = \frac{1}{2}\delta_{jk}. \\
		\bm\gamma_j' \cdot \bm\gamma_k' &= (\sin(\phi) \bm\beta_j + \cos(\phi) \bm\gamma_j) \cdot (\sin(\phi) \bm\beta_k + \cos(\phi) \bm\gamma_k), \nonumber\\
		&= (\sin^2(\phi) \bm\beta_j\cdot \bm\beta_k + \sin(\phi) \cos(\phi) \bm\beta_j \cdot \bm\gamma_k + \sin(\phi) \cos(\phi) \bm\gamma_j \cdot \bm\beta_k + \cos^2(\phi) \bm\gamma_j \cdot \bm\gamma_k) = \frac{1}{2}\delta_{jk}.
	\end{align}
	with \(j, k = 1, \ldots, d-1\). We can see that the new coefficients $\{\beta'^\alpha_0,\beta'^\alpha_j,\gamma'^\alpha_j\}_{j=1}^{d-1}$ of the new POVM satisfy the conditions to be a FSM, Eq.\thinspace\eqref{FSMconditions}, therefore $\mathcal{E}'$ is also a FSM.

    \subsubsection{ Collections of FSMs are also FSMs}\label{apendix4}
	
	In this section, we demonstrate that a collection of $m$ FSMs $\{ \mathcal{E}_i \}_{i=0}^{m-1}$, with $\mathcal{E}_i = \{ \ket*{\varphi_i^{\alpha}}\}_{\alpha=0}^{n-1}$, $n \geq 2d-1$, $\alpha=0,...,n-1$, and
	\begin{equation}
		\ket*{\varphi_i^{\alpha}} = \beta_{i,0}^{\alpha}\ket*{0} + \sum_{k=1}^{d-1}(\beta_{i,k}^{\alpha}+i\gamma_{i,k}^{\alpha})\ket*{k}, \quad \beta_{i,0}^l > 0, \quad \beta_{i,k}^{\alpha},\gamma_{i,k}^{\alpha}\in \mathbb{R},
	\end{equation}
	is also an FSM. The coefficients $\{  \beta_{i,0}^{\alpha}, \beta_{i,k}^{\alpha}, \gamma_{i,k}^{\alpha}) \}$ satisfy the FSM conditions, Eq.\thinspace\eqref{FSMconditions}, for each $i=0,\dots, m-1$. \\
	
	Consider POVM $\mathcal{E}' = \{\tau_i^2\ket*{\varphi_i^{\alpha}}\bra*{\varphi_i^{\alpha}}\}$, with positive coefficients $\{\tau_i\}$ such as $\sum_i|\tau_i|^2=1$. Defining the coefficient of each POVM element,
	\begin{align}
		\beta'^\alpha_{i,0} =& \tau_i\beta^\alpha_{i,0},	\\
		\beta'^\alpha_{i,k} =& \tau_i\beta^\alpha_{i,k},	\\
		\gamma'^\alpha_{i,k} =& \tau_i\gamma^\alpha_{i,k},	
	\end{align}
	and their respective $mn-$dimensional vectors $\bm\beta'_0=(\beta'^0_{0,0},\dots, \beta'^n_{m,0})$, $\bm\beta'_k=(\beta'^0_{0,k},\dots, \beta'^n_{m,k})$,  $\bm\gamma'_0=(\gamma'^0_{0,k},\dots, \gamma'^n_{m,k})$, we can verify if $\mathcal{E}'$ satisfies the conditions for being an FSM, Eq.\thinspace\eqref{FSMconditions}, 
	\begin{align}
		\bm\beta'_0\cdot \bm\beta'_0 &= \sum_{i=0}^{m-1}|\tau_i|^2 \sum_{\alpha=0}^{n_i-1} \beta_{i,0}^{\alpha} \beta_{i,0}^{\alpha} = 1,\\
		\bm\beta'_0\cdot \bm\beta'_k &= \sum_{i=0}^{m-1}|\tau_i|^2 \sum_{\alpha=0}^{n_i-1} \beta_{i,0}^{\alpha} \beta_{i,k}^{\alpha}=0,\\
		\bm\beta'_0\cdot \bm\gamma'_k &= \sum_{i=0}^{m-1}|\tau_i|^2 \sum_{\alpha=0}^{n_i-1} \beta_{i,0}^{\alpha} \gamma_{i,k}^{\alpha}=0,\\
		\bm\beta'_{j}\cdot \bm\gamma'_{k} &= \sum_{i=0}^{m-1}|\tau_i|^2 \sum_{\alpha=0}^{n_i-1} \beta_{i,j}^{\alpha} \gamma_{i,k}^{\alpha} =0,\\
		\bm\beta'_{j}\cdot \bm\beta'_{k} &= \sum_{i=0}^{m-1}|\tau_i|^2 \sum_{\alpha=0}^{n_i-1} \beta_{i,j}^{\alpha} \beta_{i,k}^{\alpha} = \frac{1}{2}\delta_{jk},\\
		\bm\gamma'_{j}\cdot \bm\gamma'_{k} &= \sum_{i=0}^{m-1}|\tau_i|^2 \sum_{\alpha=0}^{n_i-1} \gamma_{i,j}^{\alpha} \gamma_{i,k}^{\alpha} =  \frac{1}{2}\delta_{jk}.
	\end{align}
	From these equations, we can see that the POVM $\mathcal{E}'$ satisfies the FSM conditions. Thus, the set $\mathcal{E}'$ corresponds to an FSM composed of $mn$ elements.
	
	%%%%%%%%%%%%%%%%%%%%%%%%%%%%%%%%%%%%%%%%%%%%%%%%%%%%%%%%%%%%%%%%%%%%%%%%
	
	%##############################################

    \subsection{Analytical reconstruction}\label{sec:analy_recon}
    In this section, we describe how to retrieve the parameters $\{ x_{j\sigma} \}$ from the probabilities $p(\alpha|x)$ Eq.\thinspace\eqref{eq:prob_fsm_opt} after measuring a FSM. 
    
    Let us consider the following linear combination of the probabilities,
    \begin{align}
        \Delta_j = \sum_\alpha \frac{\omega_j^\alpha}{\beta_0^\alpha} p_\alpha,
    \end{align}
    with $\omega_j^\alpha=\beta_j^\alpha + i\gamma_j^\alpha$ the coefficients of the FSM. Using the conditions on a POVM to be an FSM, Eqs.\thinspace\eqref{eq:cond1} and \eqref{eq:cond2}, we obtain the following non-linear system of equations 
    \begin{align}
        \Delta_j = a_0(x_{j0} + i x_{j1}). \label{eq:fsm_opt_lin_coef}
    \end{align}
    Combining this equation with the normalization condition, we obtain an equation only for the parameter $a_0$,
    \begin{align}
        \sum_{j=1}^{d-1} |\Delta_j|^2 = a_0^2 \sum_{j=1}^{d-1}\sum_{\sigma=0}^1(x_{j\sigma})^2 = a_0^2(1-a_0^2).
    \end{align}
    Suppose that $a_0\approx1$, we can write $a_0^2=1-\xi^2$, with $\xi\approx0$. Then
    \begin{align}
        \sum_{j=1}^{d-1}  |\Delta_j|^2 = (1-\xi^2)\xi^2 \approx \xi^2. \label{eq:cal_norm_phi}
    \end{align}
    Therefore, we find that $a_0^2$ is approximately given by
    \begin{align}
        a_0^2\approx 1- \sum_{j=1}^{d-1}  |\Delta_j|^2.
    \end{align}
    Changing $a_0^2$ in Eq.\thinspace\eqref{eq:fsm_opt_lin_coef}, we are able to isolate all parameters as
    \begin{align}
        x_{j0} = {\rm Re}\left(  \frac{\Delta_j}{\sqrt{1- \sum_{j=1}^{d-1}  |\Delta_j|^2}}\right), \qquad x_{j1} = {\rm Im}\left(  \frac{\Delta_j}{\sqrt{1- \sum_{j=1}^{d-1}  |\Delta_j|^2}}\right). \label{eq:fsm_opt_sol_tomo}
    \end{align}
    
    \subsection{Error analysis}
    The FSM allows us to obtain an estimator that saturates the Gill-Massar inequality and achieves a mean infidelity of $\bar{I}=(d-1)/N$. However, this result is an average across all possible estimators, so individual estimators may perform better or worse. Additionally, the infidelity of the estimator deteriorates as the overlap with the fiducial, denoted by \(a_0\), decreases. Due to these two reasons, the infidelity in the estimation may deviate from the Gill-Massar bound. In this section, we derive a high-probability upper bound on infidelity in estimation with FSM, expressed in terms of the sample size \(N\) and the overlap \(a_0\) with the fiducial state.

    \subsubsection{Linear estimator}
    Recalling Eq.\thinspace\eqref{eq:fsm_opt_lin_coef}, we define the $(d-1)$-dimensional vector
    \begin{align}
        \ket*{\Phi} = \sum_{j=1}^{d-1} \Delta_j \ket*{j}.
    \end{align}
    Note that the vector $\ket*{\Phi}$ is different from the state $\ket*{\Psi}$. Furthermore, its 2-norm is
    \begin{align}
        ||\ket*{\Phi}||_2 = \sqrt{\sum_{j=1}^{d-1} |\Delta_j|^2} = \sqrt{a_0^2(1-a_0^2)} \leq \frac{1}{2},
    \end{align}
    so it is not a normalized vector, then it is not a quantum state either. However, it has the advantage of being expressed as a linear estimator of the probabilities,
    \begin{align}
        \ket*{\Phi} = \sum_{\alpha=1}^{n} p_\alpha\ket*{\phi_\alpha},
    \end{align}
    where $\{\ket*{\phi_\alpha} \}$ is a set of known vectors, depending only on the FSM, given by
    \begin{align}
        \ket*{\phi_\alpha} = \sum_{j=1}^{d-1} \frac{\omega_j^\alpha}{\beta_0^\alpha} \ket*{j}.
    \end{align}
    Moreover, we can interpret $\ket*{\Phi}$ as the expected value of the vector-valued random variable $\{\ket*{\phi_\alpha} \}$, that is,
    \begin{align}
        \ket*{\Phi} = \mathbb{E}(\ket*{\phi_\alpha}).
    \end{align}
    Note that, according to Eq.\thinspace\eqref{eq:fsm_opt_sol_tomo}, estimating \(\ket*{\Phi}\) is equivalent to \(\ket*{\Psi}\). This equivalence allows us to upper bound the error in estimating \(\ket*{\Psi}\) by the error in estimating \(\ket*{\Phi}\). 
    
    \subsubsection{Upper-bound of error in 2-norm}
    Let \(|\hat{\Psi}\rangle \) be an estimate of \(\ket*{\Psi}\). We will consider the 2-norm distance between \(\ket*{\Psi}\) and \(|\hat{\Psi}\rangle \) as the error metric. Defining $\ket*{\Psi_1} = \sum_{j=1}^{d-1} (x_{j0}+ix_{j1}) \ket*{j}$, we have that
    \begin{align}
        |\ket*{\Psi} - |\hat{\Psi}\rangle|_2 \le |a_0 - \hat{a}_0| + |\ket*{\Psi_1} - |\hat{\Psi}_1\rangle|_2,
    \end{align}
    where $\hat{a}_0$ and $|\hat\Psi_1\rangle$ are estimators of $a_0$ and $\ket*{\Psi_1}$, respectively. Scaling by $a_0$, rearranging terms, and applying the triangular inequality, we find
    \begin{align}
	|\ket*{\Psi}-|\hat{\Psi}\rangle|_2 \leq& | a_0 -\hat{a}_0| + \frac{1}{a_0}|a_0\ket*{\Psi_1} - a_0|\hat\Psi_1\rangle   |_2\\
	=& | a_0 -\hat{a}_0| + \frac{1}{a_0}|a_0\ket*{\Psi_1} +\hat{a}_0|\hat\Psi_1\rangle-\hat{a}_0|\hat\Psi_1\rangle - a_0|\hat\Psi_1\rangle   |_2 \\
	\leq& | a_0 -\hat{a}_0| + \frac{1}{a_0}\left( |a_0\ket*{\Psi_1} - \hat{a}_0|\hat\Psi_1\rangle |_2 + |a_0-\hat{a}_0| \right)\\
	=& \left(1+\frac{1}{a_0}\right)| a_0 -\hat{a}_0| + \frac{1}{a_0} |\ket*{\Phi}-\ket*{\hat\Phi}|_2 , \label{eq:bound_error_1}
    \end{align}
where $|\hat\Phi\rangle$ is an estimator of $\ket*{\Phi}$. Thus, the error in estimating the state $\ket*{\Psi}$ can be upper bounded in terms of the errors in estimating the coefficient $a_0$ and the vector $\ket*{\Phi}$. Given that the coefficient $a_0$ can be obtained from $\ket*{\Phi}$, their errors can also be related. Recalling that we can write $ a_0^2 \approx 1 - \xi^2$, it follows that
\begin{align}
	|a_0 - \hat{a}_0| &= \left| \sqrt{1 - \xi^2} - \sqrt{1 - \hat\xi^2} \right| \\
	&\approx \left| (1 - \xi^2/2) - (1 - \hat\xi^2/2) \right| \\
	&= \frac{1}{2} |\xi^2 - \hat\xi^2|,
\end{align}
with $\hat\xi^2$ the estimation of $\xi^2$. Noticing from Eq.\thinspace\eqref{eq:cal_norm_phi} that $\xi^2 = \braket*{\Phi}{\Phi}$ and that $|\ket*{\Phi}|_2\leq 1/2$, we can upper bound the error in estimating $\xi^2$ as follows,
\begin{align}
	|\xi^2 - \hat\xi^2| = | \braket*{\Phi}{\Phi} - \braket*{\hat\Phi}{\hat\Phi} | \leq (2 |\ket*{\Phi}|_2+|\ket*{\Phi} - |\hat\Phi\rangle|_2)\,|\ket*{\Phi} - |\hat\Phi\rangle|_2 \lesssim |\ket*{\Phi} - |\hat\Phi\rangle|_2. \label{eq:upper_boud_norm_phi}
\end{align}
Then, to first order in $\xi^2$ and in $|\ket*{\Phi} - |\hat\Phi\rangle|_2$, the error in estimating $a_0$ is bounded by the error in estimating $\ket*{\Phi}$,
\begin{align}
	|a_0 - \hat{a}_0| \lesssim \frac{1}{2}|\ket*{\Phi} -\ket*{\hat\Phi}|_2.
\end{align}
Using the previous equation in Eq.\thinspace\eqref{eq:bound_error_1}, we obtain the following.
\begin{align}
	|\ket*{\Psi} - |\hat{\Psi}\rangle|_2 \lesssim \frac{2}{a_0} |\ket*{\Phi} - |\hat{\Phi}\rangle|_2. \label{eq:upper_bound_error_1}
\end{align}
In conclusion, the error in estimating the state $\ket*{\Psi}$ is upper-bounded by the error in estimating $\ket*{\Phi}$. Consequently, by constructing a concentration inequality for $\ket*{\Phi}$, we obtain a concentration inequality for the full state $\ket*{\Psi}$.

\subsubsection{Confidence region}\label{sec:conf_region}
Confidence regions provide guarantees of the accuracy of a tomographic method as a function of the sample size $N$. These arise from interpreting the tomographic estimator as a sum of matrix-valued random variables and applying concentration inequalities \cite{Guta2020}. In our case $\ket*{\Phi}$ can be interpreted as a sum of vector-valued random variables $\ket*{\phi_\alpha}$, so we can employ the following concentration inequality \cite{deGois2024}:\\

\fbox{\parbox{.95\textwidth}{Let $\ket*{X_1}, \dots, \ket*{X_N}$ be $N$ independent, zero-mean, vector-valued random variables. Then, the following concentration inequality holds for any $\epsilon>0$:
\begin{align}
	\mathbb{P}\left[ \left| \frac{1}{N} \sum_{i=1}^N \ket*{X_i} \right|_2 \ge \epsilon \right] \le 8 \exp\left( - \frac{N\epsilon^2}{2\sigma^2} w\left( \epsilon \frac{L}{\sigma} \right) \right), \label{eq:concentration_inequality}
\end{align}
where $w(x) = \frac{3}{3+x}$ and
\begin{align}
    L \ge|\ket*{X_i}|_2, \qquad \sigma^2 \ge \sum_{i=1}^N\frac{1}{N}\mathbb{E}[|\ket*{X_i}|_2^2].
\end{align}}}\\

We can apply this concentration inequality to the FSM case by setting $\ket*{X_i} = \ket*{\Phi} - \ket*{\phi_{\alpha_i}}$. The next step is to calculate the bounds $L$ and $\sigma^2$. To obtain the bound $L$, notice that
\begin{align}
	|\ket*{\Phi} - \ket*{\phi_\alpha}|_2 \le |\ket*{\Phi}|_2 + |\ket*{\phi_\alpha}|_2 \le \frac{1}{2} + |\ket*{\phi_\alpha}|_2.
\end{align}
Using Eq.\thinspace\eqref{eq:bound_norm_fsm_elem}, we have
\begin{align}
	|\ket*{\phi_\alpha}|_2 = \frac{1}{\beta_0^\alpha} \sqrt{ \sum_j |\omega_j^\alpha|^2 } \le \frac{1}{\sqrt{2}}\max_\alpha \left( \frac{1}{\beta_0^\alpha} \right).
\end{align}
Thus,
\begin{align}
	|\ket*{\Phi} - \ket*{\phi_\alpha}|_2 &\le \frac{1}{2} + \frac{1}{\sqrt{2}}\max_\alpha \left( \frac{1}{\beta_0^\alpha} \right)\\
    &\leq \left( \frac{1}{2} + \frac{1}{\sqrt{2}} \right) \max_\alpha \left( \frac{1}{\beta_0^\alpha} \right) \\
    &< \sqrt{2}\max_\alpha \left( \frac{1}{\beta_0^\alpha} \right)\equiv L.
\end{align}
For the bound $\sigma^2$, it follows
\begin{align}
	\frac{1}{N}\sum_{i=1}^N\mathbb{E}\left(| \ket*{\Phi} - \ket*{\phi_{\alpha_i}} |_2^2\right) =& \frac{1}{N}\sum_{i=1}^N\mathbb{E}\left( |\ket*{\Phi}|_2^2 + |\ket*{\phi_{\alpha_i}}|^2_2 - 2{\rm Re}(\braket*{\Phi}{\phi_{\alpha_i}}) \right)\\
	=&  \frac{1}{N}\sum_{i=1}^N \mathbb{E}\left( |\ket*{\phi_{\alpha_i}}|_2^2 \right) - |\ket*{\Phi}|_2^2\\
	\leq& \frac{1}{2}\max_\alpha \left(\frac{1}{\beta_0^\alpha}\right)^2\equiv\sigma^2.
\end{align}
Defining the constant
\begin{align}
	\Omega = \max_\alpha \left( \frac{1}{\beta_0^\alpha} \right),
\end{align}
we conclude that
\begin{align}
	L = \sqrt{2}\Omega,\qquad \sigma^2=\frac{\Omega^2}{2}. \label{eq:L_and_sigma}
\end{align}
Notice that the coefficient $\Omega$ is related to the number of outcomes $n$ of the FSM by $\Omega^2\geq n$, so the bounds $L$ and $\sigma^2$ are tighter for a minimal FSM with $n=2d-1$. Replacing Eq.\thinspace\eqref{eq:L_and_sigma}, the concentration inequality becomes
\begin{align}
    \mathbb{P}\left[ | \ket*{\Phi} - \ket*{\hat\Phi}|_2 \ge \epsilon \right] \le 8 \exp\left( - \frac{N\epsilon^2}{\Omega^2}\,\frac{3}{3+2\epsilon} \right),
\end{align}
where we replace $\ket*{\hat\Phi}=\frac{1}{N} \sum_{i=1}^N\ket*{\phi_{\alpha_i}}$ and $w(2\epsilon)=3/(3+2\epsilon)$. We invert the inequality exactly. Setting the right-hand side equal to $\delta$ leads to the quadratic equation $3N\epsilon^2-2\Omega^2\log(8/\delta)\,\epsilon-3\Omega^2\log(8/\delta)=0$, whose positive root satisfies $\epsilon\leq\Omega\sqrt{\log(8/\delta)/N}+2\Omega^2\log(8/\delta)/(3N)$. Thereby, with probability at least $1-\delta$,
\begin{align}
    | \ket*{\Phi} - \ket*{\hat\Phi}|_2 \leq \epsilon_\Phi, \label{eq:phi_err}
\end{align}
with
\begin{align}
    \epsilon_\Phi = \Omega\sqrt{\frac{\log(8/\delta)}{N}}\left(1+\frac{2}{3}\,\Omega\sqrt{\frac{\log(8/\delta)}{N}}\right). \label{eq:eps_phi}
\end{align}
This expression holds for any $\epsilon$. Its leading term coincides with the approximation $w\approx1$, and the correction is of relative order $\Omega\sqrt{\log(8/\delta)/N}$, which vanishes for large $N$. Taking into account Eq.\thinspace\eqref{eq:upper_bound_error_1}, which holds to first order in the estimation error, the error $\epsilon=|\ket*{\Psi} - |\hat\Psi\rangle|_2$ in the state satisfies $\epsilon^2\lesssim4\epsilon_\Phi^2/a_0^2$. Replacing Eq.\thinspace\eqref{eq:eps_phi}, we obtain, with probability at least $1-\delta$,
\begin{align}
	\epsilon^2 \lesssim \frac{4\Omega^2}{a_0^2N} \log\left( \frac{8}{\delta} \right)\left(1+\frac{2}{3}\,\Omega\sqrt{\frac{\log(8/\delta)}{N}}\right)^2. \label{eq:error_2_norm}
\end{align}

\subsubsection{Upper-bound for infidelity}

In the previous section, we obtained an upper bound for the error in the 2-norm, which is not directly comparable to the Gill-Massar bound, which is a lower bound for infidelity. Due to that, we will use Eq.\thinspace\eqref{eq:error_2_norm} to obtain an upper bound for infidelity. For normalized states, the error in 2-norm satisfies
\begin{align}
	|\ket*{\Psi}-|\hat\Psi\rangle|_2^2 = 2\left( 1- {\rm Re}(\braket*{\Psi}{\hat\Psi}) \right) \geq 2( 1 - |\braket*{\Psi}{\hat\Psi}|).
\end{align}
Isolating the fidelity $|\braket*{\Psi}{\hat\Psi}|^2 $, we obtain
\begin{align}
	|\braket*{\Psi}{\hat\Psi}|^2 \geq \left( 1-\frac{1}{2}|\ket*{\Psi}-|\hat\Psi\rangle|_2^2  \right)^2 \geq 1 - |\ket*{\Psi}-|\hat\Psi\rangle|_2^2.
\end{align}
Thus, the 2-norm error upper bounds the infidelity,
\begin{align}
    I(\Psi,\hat\Psi ) = 1 - |\braket*{\Psi}{\hat\Psi}|^2 \leq |\ket*{\Psi}-|\hat\Psi\rangle|^2_2. 
\end{align}
Therefore, with probability at least $1-\delta$, the infidelity is upper bounded by
\begin{align}
    I(\Psi,\hat\Psi ) \lesssim \frac{4\Omega^2}{a_0^2N} \log\left( \frac{8}{\delta} \right)\left(1+\frac{2}{3}\,\Omega\sqrt{\frac{\log(8/\delta)}{N}}\right)^2. 
\end{align}
Considering the particular case of the explicit FSM introduced in section \ref{sec:exp_cons_fsm}, we have $\Omega^2=n=2d-1$. Therefore, the upper bound for infidelity becomes
\begin{align}
	I(\Psi,\hat\Psi ) \lesssim \frac{4(2d-1)}{a_0^2N} \log\left( \frac{8}{\delta} \right)\left(1+\frac{2}{3}\sqrt{\frac{(2d-1)\log(8/\delta)}{N}}\right)^2. \label{eq:upper_bound_inf}
\end{align}
This indicates that the infidelity scales as $\mathcal{O}(d/a_0^2N)$, with a leading prefactor $8\log(8/\delta)$ times larger than the one of the Gill-Massar bound. Furthermore, the upper bound increases as the fiducial overlap $a_0$ decreases. This degradation does not represent an inherent limitation of Fisher-symmetric measurements, but it comes from the conservativeness of the analytical bound, which combines a norm inequality with a worst-case concentration estimate. Therefore, the expression above should be considered a worst-case upper bound, valid to first order in the estimation error. In contrast, the Gill-Massar bound is asymptotically tight and provides a lower bound on the achievable mean infidelity.

\subsubsection{Error bound due to approximation}\label{sec:error_aprox}

The previous calculations were performed by approximating the probabilities up to first order in the parameters. Nevertheless, from a real experiment, we measure the actual probabilities without approximation, which introduces a systematic error in the estimation \cite{FSM_EXP}. The exact density matrix can be written as
\begin{align}
    \tilde\rho = \rho_1 + (1-a_0^2)\sigma,
\end{align}
where $\rho_1 = a_0^2\ket*{0}\!\bra*{0} + a_0\sum_{j=1}^{d-1}\left(z_j\ket*{j}\!\bra*{0} + z_j^*\ket*{0}\!\bra*{j}\right)$, with $z_j=x_{j0}+ix_{j1}$, contains the terms of order zero and one in $\{z_j\}$, for which Eq.\thinspace\eqref{eq:fsm_opt_lin_coef} holds exactly, and $\sigma$ is the density matrix
\begin{align}
    \sigma = \frac{1}{1-a_0^2} \sum_{j,k=1}^{d-1} z_j z_k^* \,|j\rangle\thinspace\langle k |.
\end{align}
Thus, the actual probability distribution is given by
\begin{align}
	\tilde{p}_\alpha = p_\alpha + (1 - a_0^2) q_\alpha,
\end{align}
where $p_\alpha={\rm Tr}(E^\alpha\rho_1)$ and $q_\alpha={\rm Tr}(E^\alpha\sigma)$ is a probability distribution. The resulting linear estimator becomes
\begin{align}
	|\tilde{\Phi}\rangle = \ket*{\Phi} + (1 - a_0^2) \ket*{Q}, 
\end{align}
with $\ket*{Q} = \sum_\alpha q_\alpha \ket*{\phi_\alpha}$. Let $|\hat{\tilde{\Phi}}\rangle$ be the estimator of $\ket*{\Phi}$ obtained from the frequencies of $\tilde{p}_\alpha$. According to the triangle inequality, its error is at most the statistical error $| |\tilde{\Phi}\rangle - |\hat{\tilde{\Phi}}\rangle |_2$, which is bounded by $\epsilon_\Phi$ of Eq.\thinspace\eqref{eq:eps_phi} with probability $1-\delta$, plus the norm $(1-a_0^2)|\ket*{Q}|_2$. Since $|\ket*{Q}|_2\leq\sum_\alpha q_\alpha |\ket*{\phi_\alpha} |_2$ and $|\ket*{\phi_\alpha} |_2\leq\Omega/\sqrt{2}$, we obtain
\begin{align}
	| \ket*{\Phi} - |\hat{\tilde{\Phi}}\rangle |_2 \leq \epsilon_\Phi + \kappa,
\end{align}
with
\begin{align}
	\kappa=(1-a_0^2) \frac{\Omega}{\sqrt{2}}.
\end{align}
The first term is the statistical error and the second term is the bias due to the approximation. Using Eq.\thinspace\eqref{eq:upper_bound_error_1}, with probability $1-\delta$,
\begin{align}
    I(\Psi, \hat{\tilde \Psi} ) \lesssim \frac{4}{a_0^2}\left(\epsilon_\Phi + \kappa\right)^2 .
\end{align}
Therefore, depending on the relationship between $\epsilon_\Phi$ and $\kappa$, the statistical error could be hidden by the bias. When $\kappa\gtrsim\epsilon_\Phi$, the bound saturates at $4\kappa^2/a_0^2$, which does not decrease with the sample size $N$. Since the bias of the linear estimator is generically of the order of $1-a_0^2$, its infidelity exhibits a floor, a behavior observed in experiments \cite{FSM_EXP}. We have a small bias floor, $\kappa\ll1$, when
\begin{align}
    a_0^2 \gg 1 - \frac{\sqrt{2}}{\Omega},
\end{align}
which for the explicit FSM of section \ref{sec:exp_cons_fsm} reads $a_0^2 \gg 1 - \sqrt{2/(2d-1)}$. However, the bias is negligible with respect to the statistical error, $\kappa\ll\epsilon_\Phi$, only when
\begin{align}
    1-a_0^2 \ll \sqrt{\frac{2\log(8/\delta)}{N}},
\end{align}
which becomes more demanding as $N$ increases. Both conditions are sufficient but not necessary, since they are derived from upper bounds. Let us note that the floor is a property of the linearized reconstruction and not of the FSM data, since a maximum likelihood estimator based on the exact probabilities does not exhibit it.

%%%%%%%%%%%%%%%%%%%%%%%%%%%%%%%%%%%%%%%%%%%%%%%%%%%%%%%%%%%%%%%%%%%%%%%%%%%%%%%%%%%%%%%%%%%%%%%%%%%%%%%%%%%%%%%%%%%%%%%%%%

    \section{Pure state estimation with two FSMs}\label{apendix3}

    In this section, we demonstrate that any pure state, except a null subset, can be estimated with a measurement into two FSMs. We then perform an error analysis of this method.

    \subsection{Analytical reconstruction}
	Consider a POVM composed of two FSMs $\mathcal{E}_\pm = \{\ket*{\varphi^\alpha_\pm}\bra*{\varphi^\alpha_\pm}\}$, where
	\begin{align}
		\ket*{\varphi^\alpha_\pm}=\beta_0^\alpha\ket*{0} \pm \sum_{k=1}^{d-1}\omega_k^\alpha \ket*{k},
	\end{align}
    with $\beta_0^\alpha$ and $\omega_k^\alpha$ satisfying the orthogonality conditions presented in section\thinspace\ref{sec:ortho_cond}, and with $\alpha = 0,...,n-1$ and $k= 1,...,d-1$. Let us consider an arbitrary state given by Eq.\thinspace\eqref{eq:state}, but without assuming that the parameters are small. This means that our state could be far from the fiducial state $\ket*{0}$. Measuring this state with the POVM composed of two FSMs, we get
	\begin{align}
		P_\pm^\alpha &= | \langle \varphi_\pm^\alpha|\Psi \rangle |^2 = (\beta_0^\alpha)^2 a_0^2 + \left|\sum_{j=1}^{d-1} (\omega_j^\alpha)^* z_j \right|^2 \pm \beta_0^\alpha a_0\sum_{j=1}^{d-1}\left[\left(\omega_j^\alpha\right)^* z_j + \omega_j^\alpha z_j^* \right],
	\label{eq:probs_2fsm}\end{align}
    where we defined $z_j=x_{j0}+ix_{j1}$ for simplicity. Subtracting these expressions,
	\begin{align}
		P_+^\alpha -P_-^\alpha = 2a_0\beta_0^\alpha \sum_{j=1}^{d-1}\left[\left(\omega_j^\alpha\right)^* z_j + \omega_j^\alpha z_j^* \right]. \label{eq:resta_probs}
	\end{align}
	Multiplying the previous expression by $\omega_k^\alpha/\beta_0^\alpha$, we have
	\begin{align}
		\left(\frac{\omega_k^\alpha}{\beta_0^\alpha}\right)(P_+^\alpha -P_-^\alpha)  &= 2a_0 \sum_{j=1}^{d-1}        \left[\omega_k^\alpha\left(\omega_j^\alpha\right)^* z_j + \omega_k^\alpha\omega_j^\alpha z_j^* \right],
	\end{align}
	Summing over all $\alpha$ and using the orthogonality conditions of Eqs.\thinspace\eqref{eq:cond1} and \eqref{eq:cond2}, we obtain
	\begin{align}
		\sum_{\alpha=1}^n \left(\frac{\omega_k^\alpha}{\beta_0^\alpha}\right)(P_+^\alpha -P_-^\alpha)  &= 2a_0 \sum_{j=1}^{d-1}        \left[\sum_{\alpha=1}^n \omega_k^\alpha\left(\omega_j^\alpha\right)^* z_j + \sum_{\alpha=1}^n \omega_k^\alpha\omega_j^\alpha z_j^* \right] = 2a_0 z_k.
	\end{align}
	This expression motivates us to define the quantities
	\begin{align}
		\Delta_k &= \frac{1}{2}\sum_{\alpha=1}^n \left(\frac{\omega_k^\alpha}{\beta_0^\alpha}\right)(P_+^\alpha -P_-^\alpha). \label{eq:delta_2fsm}
    \end{align}
	The quantities $\Delta_k$ can be measured experimentally, so we can use them to find the coefficients of the unknown pure state $\ket*{\Psi}$ as \begin{align}
	    \Delta_k=a_0 z_k. \label{eq:2FSM_delta_params}
	\end{align} 
    To isolate the parameters $z_k$ from Eq.\thinspace\eqref{eq:2FSM_delta_params}, we have to find the parameter $a_0$. This can be done by noticing that
    \begin{align}
		\sum_{k=1}^{d-1}\left| \Delta_k \right|^2 = a_0^2(1-a_0^2), \label{eq:quadratic_1}
    \end{align}
    so, we have a quadratic equation for $a_0^2$, whose solutions are
    \begin{align}
        a_0^2 = \frac{1}{2}\left( 1\pm\sqrt{1-4\sum_{k=1}^{d-1}\left| \Delta_k \right|^2} \right)    . \label{eq:sol_quadratic_1}
    \end{align}
    We have two possible solutions for \(a_0^2\), but only one of them is valid. To determine the correct solution, notice that when we add the probabilities \(P_\pm^\alpha\), we find
	\begin{equation}
		P_+^\alpha + P_-^\alpha = 2(\beta_0^\alpha)^2 a_0^2 + 2\left|\sum_{k=1}^{d-1} \left(\omega_k^\alpha\right)^* z_k \right|^2.
	\end{equation}
	Multiplying the previous equation by $a_0^2$, we obtain
	\begin{equation}
		a_0^2 (P_+^\alpha + P_-^\alpha) = 2(\beta_0^\alpha)^2 (a_0^2)^2 + 2(a_0)^2\left|\sum_{k=1}^{d-1} \left(\omega_k^\alpha\right)^* z_k \right|^2.
	\end{equation}
    Recalling Eqs.\thinspace\eqref{eq:2FSM_delta_params} and \thinspace\eqref{eq:quadratic_1}, we find a set of $n$ linear equations for $a_0^2$,
	\begin{align}
		a_0^2 (P_+^\alpha + P_-^\alpha) &= 2(\beta_0^\alpha)^2 \left(a_0^2 - \sum_{k=1}^{d-1}|\Delta_k|^2\right) + 2 \left|\sum_{k=1}^{d-1} \left(\omega_k^\alpha \right)^*\Delta_k\right|^2.
	\end{align}
	Solving these equations, we obtain $n$ estimators of $a_0$,
	\begin{align}
		a_0^\alpha & = \sqrt{\frac{ \left|\sum_{k=1}^{d-1} \left(\omega_k^\alpha\right)^*\Delta_k\right|^2  -(\beta_0^\alpha)^2\sum_{k=1}^{d-1}|\Delta_k|^2}{(P_+^\alpha+P_-^\alpha)/2 -(\beta_0^\alpha)^2}}. \label{eq:sol_quad_2}
	\end{align}
    Using these estimators of $a_0$ we can discriminate which is the correct solution in Eq.\thinspace\eqref{eq:sol_quadratic_1}, or take the median of the estimators with $0<(a_0^\alpha)^2<1$.
    Having $a_0$, we can determine the coefficients \( z_k \) by substituting \( a_0 \) into Eq.~\eqref{eq:2FSM_delta_params}. However, this method fails when \( a_0 = 0 \), as Eq.~\eqref{eq:resta_probs} would become null for all values of \( \alpha \). Therefore, with measurements from two FSMs, we can identify the parameters of any arbitrary pure state, except for the null subspace where \( a_0 = 0 \).
	
    \subsection{Error Analysis}

    \subsubsection{Linear Estimator}
    Similarly to the optimal FSM case, we can define a linear estimator for the vector
    \begin{align}
        \ket*{\Phi} = \sum_{j=1}^{d-1}\Delta_j\ket*{j}
    \end{align}
    Replacing Eq.\thinspace\eqref{eq:delta_2fsm}, we obtain
    \begin{align}
	\ket*{\Phi} =& \sum_{\alpha=1}^n\sum_{s=\pm} ( s P_s^\alpha ) \sum_{k=1}^{d-1}\frac{1}{2}\frac{\omega_k^\alpha}{\beta_0^\alpha}\ket*{k}
\end{align}
Defining the random vectors
\begin{align}
	\ket*{\Phi_\pm^\alpha} = \pm \sum_{k=1}^{d-1}\frac{\omega_k^\alpha}{\beta_0^\alpha}\ket*{k},
\end{align}
we find that $\ket*{\Phi}$ is given by the expected value of $\{\ket*{\Phi_\pm^\alpha} \}$ on the probability distribution $\{P^\alpha_\pm/2\}$, that is,
\begin{align}
	\ket*{\Phi} = \mathbb{E}(\ket*{\Phi_\pm^\alpha}).
\end{align}

\subsubsection{Error upper bound}
The inequality given by Eq.\thinspace\eqref{eq:bound_error_1} also holds for this case, so
\begin{align}
	|\ket*{\Psi}-|\hat{\Psi}\rangle|_2 \leq\left(1+\frac{1}{a_0}\right)| a_0 -\hat{a}_0| + \frac{1}{a_0} |\ket*{\Phi}-\ket*{\hat\Phi}|_2. \label{eq:upper_bound_2fsm}
\end{align}
Using the concentration inequality presented in section \ref{sec:conf_region}, we have the same result as before, that is $L=\sqrt{2}\Omega$ and $\sigma^2=\Omega^2/2$. Therefore, 
\begin{align}
    \delta=\mathbb{P}\left[ | \ket*{\Phi} - \ket*{\hat\Phi}|_2 \ge \epsilon \right] \le 8 \exp\left( - \frac{N\epsilon^2}{\Omega^2}\,\frac{3}{3+2\epsilon} \right).   \label{eq:concentration_inequality_non_optimal_FSM}
\end{align}
This inequality establishes that $| \ket*{\Phi} - \ket*{\hat\Phi}|_2\le\epsilon$ with probability $1-\delta$, so we only have to deal with the upper bound of $ |a_0 -\hat{a}_0|$.

The upper bound for the error of $a_0$ can be obtained using error propagation in Eq.\thinspace\eqref{eq:quadratic_1}. Taking into account the definition of $\ket*{\Phi}$, this equation can be expressed as
\begin{align}
    a_0^2(1-a_0^2) = \braket*{\Phi}{\Phi} .
\end{align}
From Eq.\thinspace\eqref{eq:upper_boud_norm_phi}, we know that
\begin{align}
    | \braket*{\Phi}{\Phi} - \braket*{\hat\Phi}{\hat\Phi}| \lesssim | \ket*{\Phi} - \ket*{\hat \Phi} |_2 \leq \epsilon.
\end{align}
In order to propagate the error, we have to derive both sides by $\xi^2=\braket*{\Phi}{\Phi}$, obtaining
\begin{align}
    2a_0(1-a_0^2)\dv{a_0}{\xi^2} + a_0^2\left(-2a_0\dv{a_0}{\xi^2}\right) = 1.
\end{align}
Isolating the derivative, we have
\begin{align}
    \dv{a_0}{\xi^2} = \frac{1}{2a_0(1-2a_0^2)}.
\end{align}
Thus, to first order in the statistical error, the error of $a_0$ is upper-bounded by
\begin{align}
    |a_0-\hat{a}_0| \lesssim \frac{| \xi^2 - \hat{\xi}^2|}{2a_0|1-2a_0^2|}\lesssim \frac{\epsilon}{2a_0|1-2a_0^2|}. \label{eq:bound_a_1}
\end{align}
This equation establishes that the error of $a_0$ estimated by Eq.\thinspace\eqref{eq:sol_quadratic_1} scales the same as $\ket*{\Phi}$, except when $a_0\approx0$ and $a_0^2\approx1/2$, where the upper bound diverges. The first divergence around $a_0\approx0$ occurs because the protocol with two FSMs requires a fiducial with non-zero overlap with the unknown state. The second divergence around $a_0^2\approx1/2$ can be eliminated using the estimators given by Eq.\thinspace\eqref{eq:sol_quad_2}. This equation can be rewritten as
\begin{align}
    (a_0^\alpha)^2 = \frac{ |\braket*{\omega^\alpha}{\Phi}|^2 - (\beta_0^\alpha)^2\braket*{\Phi}{\Phi}}{(P_+^\alpha+P_-^\alpha)/2-(\beta_0^\alpha)^2},
\end{align}
with $\ket*{\omega^\alpha}=\sum_{k=1}^{d-1}\omega_k^\alpha\ket*{k}$. Let us define
\begin{align}
    A^\alpha =& |\braket*{\omega^\alpha}{\Phi}|^2 - (\beta_0^\alpha)^2\braket*{\Phi}{\Phi}, \\
	B^\alpha =& (P_+^\alpha+P_-^\alpha)/2-(\beta_0^\alpha)^2,
\end{align}
so that $(a_0^\alpha)^2 = A^\alpha/B^\alpha$. This permits us to upper bound the error of $a_0^\alpha$ as
\begin{align}
    |a_0^2 - (\hat{a}_0^\alpha)^2| = \left|\frac{A^\alpha}{B^\alpha}-\frac{\hat{A}^\alpha}{\hat{B}^\alpha}\right|\leq \frac{1}{|B^\alpha|}|A^\alpha-\hat{A}^\alpha| + \frac{A^\alpha}{(B^\alpha)^2}|B^\alpha-\hat{B}^\alpha| .
\end{align}
Therefore, we need to upper bound the errors for $A^\alpha$ and $B^\alpha$. For $A^\alpha$ we have
\begin{align}
	|A^\alpha-\hat{A}^\alpha| =& \left| |\braket*{\omega^\alpha}{\Phi}|^2 -(\beta_0^\alpha)^2\braket*{\Phi}{\Phi} - |\braket*{\omega^\alpha}{\hat\Phi}|^2+(\beta_0^\alpha)^2\braket*{\hat\Phi}{\hat\Phi} \right| \\
	\leq& \left| |\braket*{\omega^\alpha}{\Phi}|^2 - |\braket*{\omega^\alpha}{\hat\Phi}|^2 \right|
	+ (\beta_0^\alpha)^2 \left| \braket*{\Phi}{\Phi} - \braket*{\hat\Phi}{\hat\Phi} \right| \\
	\leq& (1 - (\beta_0^\alpha)^2 )| \ket*{\Phi}-\ket*{\hat\Phi} | + (\beta_0^\alpha)^2 | \ket*{\Phi}-\ket*{\hat\Phi} | \\
	=& | \ket*{\Phi}-\ket*{\hat\Phi} |.
\end{align}
So, the concentration inequality for $\ket*{\Phi}$ also works for $A^\alpha$, so $|A^\alpha-\hat{A}^\alpha|\leq\epsilon$.

To calculate the error upper-bound for $B^\alpha$, we apply the Bretagnolle-Huber-Carol inequality \cite{Tomomi2022, Mikosch1996WeakCA} to the multinomial random variable $P^\alpha=(P_+^\alpha+P_-^\alpha)/2$. It reads
\begin{align}
	\delta_0= \mathbb{P}\left[\sum_{\alpha=1}^n|P^\alpha-\hat{P}^\alpha| \geq \epsilon_0 \right] \leq 2^n \exp( -N \epsilon_0^2/2 ).
\end{align}
Let us define error per probability $P^\alpha$ as $\epsilon_0^\alpha=|P^\alpha-\hat{P}^\alpha|$, so that
\begin{align}
    \epsilon_0 = \sum_{\alpha=1}^n \epsilon_0^\alpha.
\end{align}
In this way, we have $|B^\alpha-\hat{B}^\alpha|=|P^\alpha-\hat{P}^\alpha|\leq \epsilon_0^\alpha$. The upper bound for the error of $(a_0^\alpha)^2$ becomes
\begin{align}
	|a_0^2 - (\hat{a}_0^\alpha)^2| \leq \frac{1}{|B^\alpha|}\left( \epsilon + a_0^2\epsilon_0^\alpha \right).
\end{align}
Using the Taylor series of square root, we can derive a bound for the error of $a_0$,
\begin{align}
	|a_0 - \hat{a}_0^\alpha| \leq \frac{1}{2a_0}\frac{\epsilon + a_0^2\epsilon_0^\alpha}{|B^\alpha|}.
\end{align}
The error becomes significant when $|B^\alpha|$ is small. Note that $B^\alpha$ is the expectation value in $\ket*{\Psi}$ of the observable
\begin{align}
\mathcal{B}^\alpha = \frac{1}{2}\left(\dyad*{\varphi^\alpha_+} + \dyad*{\varphi^\alpha_-}\right) - (\beta_0^\alpha)^2 \mathbb{I} = \dyad*{\omega^\alpha} - (\beta_0^\alpha)^2 \sum_{j=1}^{d-1}\dyad*{j}, \label{eq:B_observable}
\end{align}
that is, $B^\alpha=|\braket*{\omega^\alpha}{\Psi}|^2-(\beta_0^\alpha)^2(1-a_0^2)$. For the minimal FSM, $\braket*{\omega^\alpha}{\omega^\alpha}=(1-(\beta_0^\alpha)^2)/2>(\beta_0^\alpha)^2$, so that $\mathcal{B}^\alpha$ is indefinite. Thereby, $B^\alpha$ vanishes on a hypersurface of states, which has zero measure but is not restricted to the kernel of $\mathcal{B}^\alpha$. Moreover, for a Haar-random state $|\braket*{\omega^\alpha}{\Psi}|^2$ is typically of the order of $(1-a_0^2)\braket*{\omega^\alpha}{\omega^\alpha}/(d-1)$, which is comparable to $(\beta_0^\alpha)^2(1-a_0^2)=(1-a_0^2)/(2d-1)$. Consequently, $B^\alpha$ changes sign from outcome to outcome and is small for a non-negligible fraction of outcomes. We therefore restrict the estimation of $a_0$ to the set $\mathcal{A}$ of outcomes with $|B^\alpha|\geq (1-a_0^2)(\beta_0^\alpha)^2/2$. In practice, this is implemented by discarding the estimates outside the interval $0<(\hat a_0^\alpha)^2<1$ and taking the median of the remaining ones. For $\alpha\in\mathcal{A}$, a first-order Taylor expansion of the square root gives $|a_0 - \hat{a}_0^\alpha| \lesssim (\epsilon + a_0^2\epsilon_0^\alpha)/(2a_0|B^\alpha|)$. Using $|B^\alpha|\geq(1-a_0^2)(\beta_0^\alpha)^2/2$ and $(\beta_0^\alpha)^2\geq1/\Omega^2$, with $\Omega = \max_\alpha (1/\beta_0^\alpha)$, we obtain
\begin{align}
	|a_0 - \hat{a}_0^\alpha| \lesssim \frac{\Omega^2}{a_0}\,\frac{\epsilon + a_0^2\epsilon_0^\alpha}{1-a_0^2}.
\end{align}

\begin{figure}
    \centering
    \includegraphics[width=\linewidth]{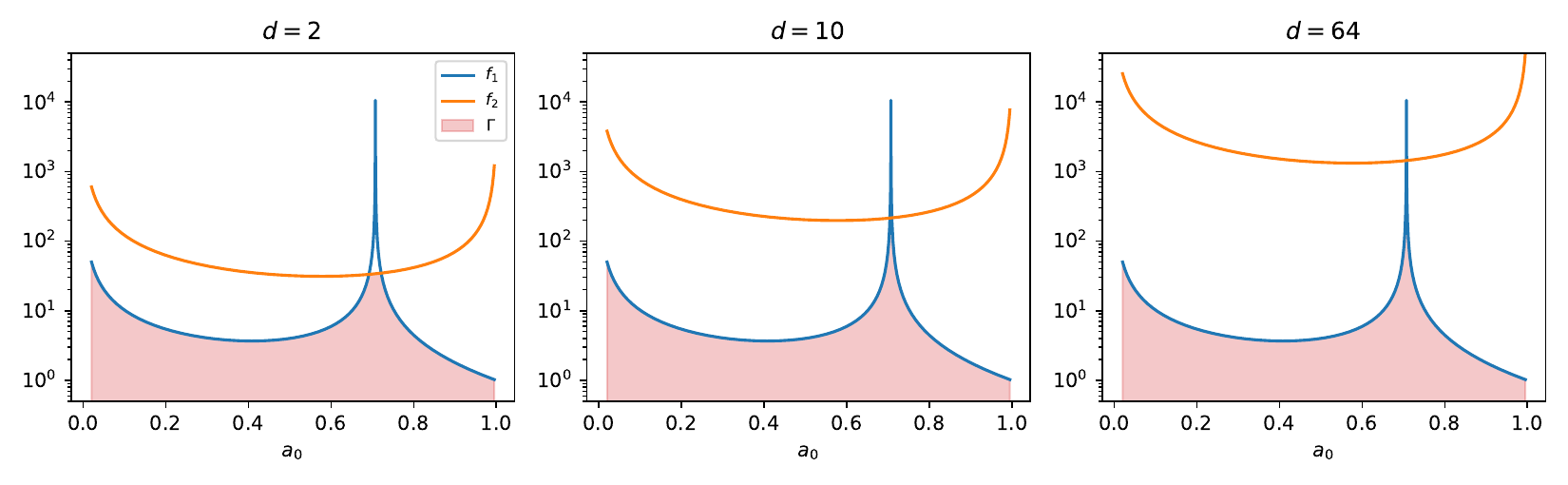}
    \caption{Functions $f_1(a_0)$ (blue line), $f_2(a_0)$ (orange line), and $\Gamma(a_0)$ (red zone) of Eq.~\eqref{eq:gamma_of_a0} for $d=2$, $10$, and $64$.}
    \label{fig:Gamma}
\end{figure}

The final estimator $\hat a_0$ is the median of $\{\hat a_0^\alpha\}_{\alpha\in\mathcal{A}}$. Since the median lies between the smallest and the largest of the individual estimates, its error is bounded by the largest individual error. Since $\epsilon_0^\alpha\leq\epsilon_0$, this leads to
\begin{align}
 |a_0-\hat{a}_0| \lesssim \frac{\Omega^2}{a_0(1-a_0^2)}\left( \epsilon + a_0^2\epsilon_0 \right). \label{eq:error_e_e0}
\end{align}
The total probability of failure ${\delta'}$ of estimating $a_0$ with this error is bounded by
\begin{align}
    {\delta'} = \delta+\delta_0 \leq 8\exp\left( -\frac{N\epsilon^2}{\Omega^2}\,\frac{3}{3+2\epsilon} \right) + 2^n \exp( -N \epsilon_0^2/2 ). 
\end{align}
Let us consider failure probabilities $\delta=\delta_0=\delta'/2$. Inverting both inequalities as in section \ref{sec:conf_region}, the errors $\epsilon$ and $\epsilon_0$ are given by
\begin{align}
    \epsilon \leq& \;\Omega\sqrt{\frac{\log(16/\delta')}{N}}\left(1+\frac{2}{3}\,\Omega\sqrt{\frac{\log(16/\delta')}{N}}\right),\\
    \epsilon_0^2 \leq& \frac{2}{N}\left[(n+1)\log 2 + \log\frac{1}{\delta'}\right].
\end{align}
Given that $\Omega^2\geq n$, we have $\epsilon\geq\epsilon_0$. Since $a_0^2\leq1$, Eq.\thinspace\eqref{eq:error_e_e0} leads to
\begin{align}
 |a_0-\hat{a}_0| \lesssim  \frac{2\Omega^2 \epsilon }{a_0(1-a_0^2)}. \label{eq:bound_a_bar_eps}
\end{align}
Compared with Eq.\thinspace\eqref{eq:bound_a_1}, this bound is worse by a factor of the order of $\Omega^2\geq n$, which scales at least as $\mathcal{O}(d)$. However, it does not blow up around $a_0^2 = 1/2$. Complementing both estimation procedures, we have the following upper bound for the error in estimating $a_0$ that only blows up around $a_0=0$,
\begin{align}
    |a_0-\hat{a}_0|\lesssim \frac{1}{2} \Gamma(a_0)\epsilon, \label{eq:final_bound_a}
\end{align}
with
\begin{align}
    \Gamma(a_0) = \min  \left( f_1(a_0), f_2(a_0)\right), \label{eq:gamma_of_a0}
\end{align}
where $f_1$ follows from Eq.\thinspace\eqref{eq:bound_a_1} and $f_2$ from Eq.\thinspace\eqref{eq:bound_a_bar_eps},
\begin{align}
    f_1(a_0)=\frac{1}{a_0|1-2a_0^2|},\qquad f_2(a_0)=\frac{4\Omega^2}{a_0(1-a_0^2)}.
\end{align}
Both functions satisfy $f_1,f_2\geq1$ for $0<a_0\leq1$. In a neighborhood of $a_0^2=1/2$, the intersection points of $f_1$ and $f_2$ are approximately given by
\begin{align}
    \left(a_{0\pm}^{\mathrm{int}}\right)^2 \approx \frac{1}{2} \pm \frac{1}{16\Omega^2},
\end{align}
and $\Gamma(a_0)=f_2(a_0)$ between these points and $\Gamma(a_0)=f_1(a_0)$ elsewhere. Figure~\ref{fig:Gamma} shows $f_1$, $f_2$, and $\Gamma$ for $d = 2$, $10$, and $64$. The maximum of $\Gamma$ is attained at $a_0^2 = 1/2$, where $\Gamma(1/\sqrt{2}) = 8\sqrt{2}\,\Omega^2$ increases linearly with $\Omega^2$, while the interval in which $\Gamma=f_2$ shrinks as $1/\Omega^2$.

Introducing Eq.\thinspace\eqref{eq:final_bound_a} into Eq.\thinspace\eqref{eq:upper_bound_2fsm} gives $|\ket*{\Psi}-\ket*{\hat{\Psi}}|_2 \lesssim (1+1/a_0)\Gamma(a_0)\epsilon/2 + \epsilon/a_0$. Since $a_0\leq1$, this is at most $[\Gamma(a_0)+1]\epsilon/a_0$, and since $\Gamma\geq1$, we obtain
\begin{align}
    |\ket*{\Psi}-\ket*{\hat{\Psi}}|_2 \lesssim \frac{2 \Gamma(a_0){\epsilon} }{a_0}.
\end{align}
with probability at least $1-\delta'$ and $\epsilon$ given above.
For the FSM of section \ref{sec:exp_cons_fsm}, $\Omega^2=n=2d-1$, so that $f_2(a_0)=4(2d-1)/[a_0(1-a_0^2)]$, $(a_{0\pm}^{\mathrm{int}})^2\approx 1/2\pm1/[16(2d-1)]$, and
\begin{align}
    I(\Psi,\hat\Psi) \lesssim \frac{ 4(2d-1)\Gamma(a_0)^2 }{ N a_0^2 }\log\left(\frac{16}{\delta'}\right)\left(1+\frac{2}{3}\sqrt{\frac{(2d-1)\log(16/\delta')}{N}}\right)^2. \label{eq:bound_inf_2fsm}
\end{align}
This bound scales as $\mathcal{O}(d/a_0^4N)$ for $a_0^2$ away from $1/2$, where $\Gamma=f_1$, and as $\mathcal{O}(d^3/N)$ for $a_0^2\approx1/2$, where $\Gamma\approx 8\sqrt{2}(2d-1)$. It is $\Gamma(a_0)^2$ times less favorable than the estimation provided by the optimal FSM, Eq.\thinspace\eqref{eq:upper_bound_inf}. This bound is conservative and holds to first order in the statistical errors; we expect the actual estimation error to be significantly smaller.

%%%%%%%%%%%%%%%%%%%%%%%%%%%%%%%%%%%%%%%%

\section{Pure state estimation with adapted FSMs}\label{sec:afsm}

In this section, we derive high-probability error bounds for the linear estimators of the adapted Fisher-symmetric measurements (AFSM). The protocol consists of three stages. In stage (0), a sample of size $n_0\approx10d$ is measured in an arbitrary basis, and the most frequent outcome is selected as the fiducial state. The largest probability $p_{\max}$ of the outcomes in this basis satisfies $p_{\max}\geq 1/d$. Thereby, its average number of counts is $n_0p_{\max}\geq10$, with fluctuations of the order of $\sqrt{n_0p_{\max}}$. Then, the most frequent outcome corresponds to a component whose overlap $a_0^2$ with the unknown state is comparable to $p_{\max}\geq1/d$. The bounds below depend on $a_0$ and hold for the fiducial state selected in this way.

In stage (1), we carry out tomography using the two non-optimal FSMs $\mathcal{E}_\pm = \{\ket*{\varphi^\alpha_\pm}\bra*{\varphi^\alpha_\pm}\}$ with a total sample size $N_1$. According to Eq.\thinspace\eqref{eq:bound_inf_2fsm}, this produces an estimate whose infidelity, with probability at least $1-\delta_1$, satisfies
\begin{align}
    \epsilon_1^2 \lesssim \frac{4(2d-1)\Gamma(a_0)^2}{N_1 a_0^2}\log\!\left(\frac{16}{\delta_1}\right)\left(1+\frac{2}{3}\sqrt{\frac{(2d-1)\log(16/\delta_1)}{N_1}}\right)^2
    \label{eq:Afsm_1}
\end{align}

In stage (2), the estimate of stage (1) is employed as fiducial state of the adapted FSM $\tilde{\mathcal{E}} = \{ U \ket*{\varphi^\alpha}\bra*{\varphi^\alpha} U^\dagger \}$, measured with a sample size $N_2$. Since the first estimate has infidelity at most $\epsilon_1^2$, the overlap $\bar{a}_0$ between the adapted fiducial state and the true state satisfies $\bar{a}_0^2 \geq 1 - \epsilon_1^2$. According to section \ref{sec:error_aprox}, the linear reconstruction of stage (2) has, with probability $1-\delta_2$, a statistical error $\epsilon_{\Phi,2}$ given by Eq.\thinspace\eqref{eq:eps_phi} with sample size $N_2$, $\Omega^2=2d-1$, and $\delta=\delta_2$, and a bias
$\kappa\leq(1-\bar{a}_0^2 )\sqrt{(2d-1)/2}$. Since $1-\bar a_0^2\leq\epsilon_1^2$,
\begin{align}
    \kappa \leq \epsilon_1^2\sqrt{\frac{2d-1}{2}}.
\end{align}
Thereby, according to section \ref{sec:error_aprox}, with probability at least $1-\delta_1-\delta_2$ the estimate of stage (2) satisfies $I(\Psi,\hat{\Psi}) \lesssim 4(\epsilon_{\Phi,2}+\kappa)^2/\bar a_0^2$, that is,
\begin{align}
    I(\Psi,\hat{\Psi}) \lesssim \left(\epsilon_2+\frac{2\kappa}{\bar a_0}\right)^2, \label{eq:final_bound_inf}
\end{align}
with $\epsilon_2=2\epsilon_{\Phi,2}/\bar a_0$, which to leading order is $\epsilon_2\approx(2/\bar a_0)\sqrt{(2d-1)\log(8/\delta_2)/N_2}$.

The adaptation provides an advantage when the bias does not dominate, $2\kappa/\bar a_0\lesssim\epsilon_2$, that is, when
\begin{align}
    \epsilon_1^2 \lesssim \sqrt{\frac{2\log(8/\delta_2)}{N_2}}.
\end{align}
In this case $I(\Psi,\hat\Psi)\lesssim 4\epsilon_2^2=\mathcal{O}(d/N_2)$, which agrees with the dimensional scaling of the Gill-Massar bound when $N_2$ is proportional to $N$. The condition requires $N_1\gtrsim d\,\Gamma(a_0)^2\sqrt{N_2}/a_0^2$, up to logarithmic factors, so that the advantage is guaranteed only for large $N$.

\begin{figure}
    \centering
    \includegraphics[width=\linewidth]{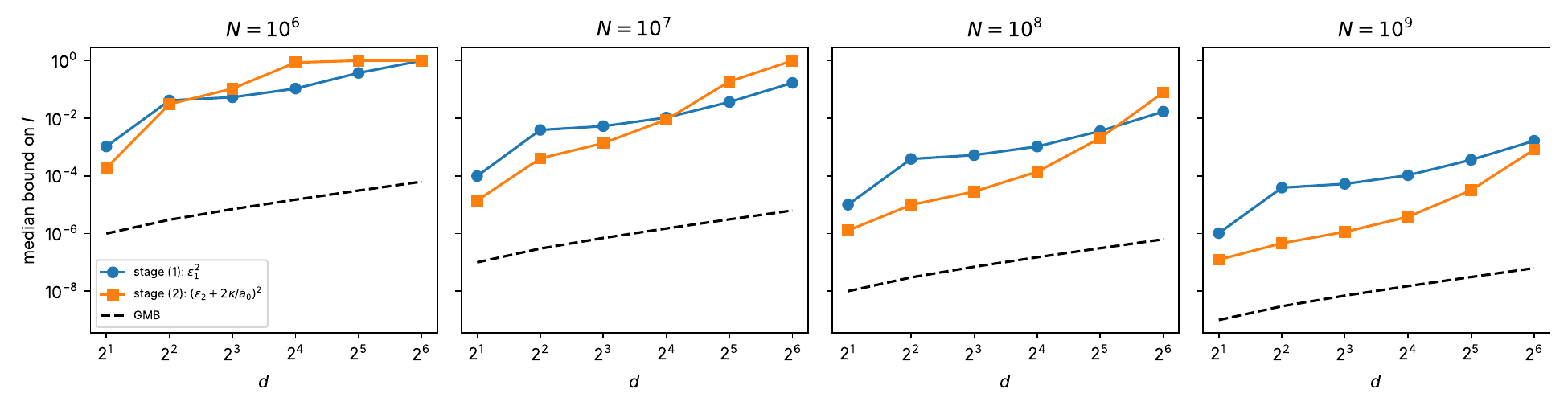}
    \caption{Median upper bounds on the infidelity of the linear estimators of stage (1), $\epsilon_1^2$ (circles), and stage (2), $(\epsilon_2+2\kappa/\bar a_0)^2$ (squares), Eqs.~\eqref{eq:Afsm_1} and \eqref{eq:final_bound_inf}, for $N_1=N_2=N/2$ with $N\in\{ 10^6,10^7,10^8,10^9 \}$ up to dimension $d=64$. Dashed lines correspond to the Gill-Massar bound $(d-1)/N$.}
    \label{fig:error_bounds}
\end{figure}

Figure~\ref{fig:error_bounds} displays the median, over $4\times10^3$ Haar-random states with the fiducial state selected from $n_0=10d$ samples, of the bounds $\epsilon_1^2$ of stage (1) and $(\epsilon_2+2\kappa/\bar a_0)^2$ of stage (2), together with the Gill-Massar bound, for $N_1 = N_2 = N/2$, $N \in \{10^6, 10^7, 10^8, 10^9\}$, and $\delta_1=\delta_2=0.05$. Bounds larger than one are set to one. The bound of stage (2) improves on the one of stage (1) for $d\leq4$, $16$, $32$, and $64$ for $N=10^6$, $10^7$, $10^8$, and $10^9$, respectively. That is, the improvement requires larger sample sizes as the dimension increases, because the bias $\kappa$ dominates unless $\epsilon_1^2$ is of the order of $N_2^{-1/2}$.

The bias $\kappa$ in Eq.~\eqref{eq:final_bound_inf} constitutes the main drawback of the present analysis. This error arises because each individual FSM is only locally informationally complete and the linear reconstruction neglects second-order terms. However, the combined measurement data from the two non-optimal FSMs $\mathcal{E}_\pm$ and the near-optimal FSM $\tilde{\mathcal{E}}$ are informationally complete for any pure state with $a_0 \neq 0$. Therefore, an estimator that jointly processes all measurement outcomes with the exact probabilities does not incur the bias $\kappa$. This motivates the use of maximum-likelihood estimation to construct the final estimator, as discussed in the next section.

\section{Attaining the Gill-Massar bound with maximum likelihood estimation}\label{sec:mle}

In this section, we show that the maximum likelihood estimator (MLE) that uses the counts of stages (1) and (2) attains the Gill-Massar bound up to a correction that decreases with the sample size. We also find the splitting of the sample size between both stages that minimizes the infidelity. We use the parametrization of Eq.\thinspace\eqref{eq:state}, where the fiducial state $\ket*{0}$ is now the fiducial state of the adapted FSM $\tilde{\mathcal{E}}$, that is, the first estimate. For large sample sizes the unknown state is close to this fiducial state. Then, we can evaluate all Fisher information matrices at the fiducial state, where $Q=4\mathbb{I}$ according to section \ref{sec:qfi}.

\subsection{Average infidelity}

Let $\bm y$ denote the MLE of the parameter vector $\bm x$. Under standard regularity conditions, the MLE of an adaptive scheme is consistent and asymptotically normal \cite{Fujiwara2006}. Thereby, for large sample sizes, $\bm y$ is approximately Gaussian distributed with mean $\bm x$ and covariance $C_N^{-1}$, where $C_N$ is the Fisher information matrix of all the data. Strictly speaking, this approximation is exact only in the limit $N\to\infty$, and deviations are expected for finite sample sizes \cite{Zhu2011,Almeida2023}. Since the samples are independent, the Fisher information of all the data is the sum of the Fisher information of each sample. Stage (1) measures $N_1/2$ samples with each FSM $\mathcal{E}_\pm$, and stage (2) measures $N_2$ samples with $\tilde{\mathcal{E}}$. We have
\begin{align}
    C_N=N_1\bar C+N_2\tilde C, \label{eq:CN}
\end{align}
where $\bar C=(C_++C_-)/2$, and $C_\pm$ and $\tilde C$ are the Fisher information matrices of $\mathcal{E}_\pm$ and $\tilde{\mathcal{E}}$. The sample size $n_0$ of stage (0) is negligible, so that $N\approx N_1+N_2$. Let us note that Eq.\thinspace\eqref{eq:CN} holds for a likelihood in which each POVM contributes with its number of counts, see Appendix~\ref{app:mle} in the main text.

In order to relate the infidelity to the error in the parameters, we use the squared Bures distance between two nearby pure states \cite{Braunstein,Paris2009},
\begin{align}
    d_B\left(\Psi(\bm x),\Psi(\bm y)\right)^2\approx\frac{1}{4}(\bm x-\bm y)^\top Q(\bm x-\bm y). \label{eq:bures_Q}
\end{align}
For pure states, $d_B^2=2(1-\sqrt F)$, with $F$ the fidelity. Since $1-F=(1-\sqrt F)(1+\sqrt F)\leq2(1-\sqrt F)$, the squared Bures distance upper bounds the infidelity. Thereby, with $Q=4\mathbb{I}$, we obtain
\begin{align}
    I(\Psi(\bm x),\Psi(\bm y))\lesssim|\bm x-\bm y|^2. \label{eq:inf_x}
\end{align}
We can see that the average infidelity is bounded by the average of $|\bm x-\bm y|^2$, that is, by ${\rm tr}\,C_N^{-1}$.

In order to evaluate ${\rm tr}\,C_N^{-1}$, we need the Fisher information matrices of the FSMs. For any rank-one POVM $\{\ket*{\phi_\alpha}\!\bra*{\phi_\alpha}\}$, Eq.\thinspace\eqref{eq:rho_approx} gives the derivatives $\partial p_\alpha/\partial x_{j0}=2{\rm Re}(\braket*{0}{\phi_\alpha}\braket*{\phi_\alpha}{j})$ and $\partial p_\alpha/\partial x_{j1}=-2{\rm Im}(\braket*{0}{\phi_\alpha}\braket*{\phi_\alpha}{j})$. Replacing them in Eq.\thinspace\eqref{Eq:CQFIM} and using $p_\alpha=|\braket*{\phi_\alpha}{0}|^2$, the trace of the Fisher information matrix becomes $4\sum_\alpha\sum_{j=1}^{d-1}|\braket*{\phi_\alpha}{j}|^2$. Since $\sum_\alpha\ket*{\phi_\alpha}\!\bra*{\phi_\alpha}=\mathbb{I}$, we obtain
\begin{align}
    {\rm tr}\,C=4(d-1). \label{eq:trC}
\end{align}
That is, every rank-one POVM saturates the Gill-Massar trace condition ${\rm tr}(Q^{-1}C)=d-1$ \cite{Gill,GillGuta2013}. We notice that an FSM distributes this trace uniformly, so that $\tilde C=2\mathbb{I}$ according to section \ref{sec:exp_cons_fsm}. In contrast, the fiducial state of $\mathcal{E}_\pm$ is far from $\ket*{\Psi}$, and $\bar C$ is not isotropic. Its eigenvalues $\lambda_i$ satisfy $\sum_i\lambda_i=4(d-1)$, that is, their average is $2$. Moreover, according to Eq.\thinspace\eqref{desigualdad_I_J}, the classical Fisher information is bounded by the quantum one, so that $0\leq\lambda_i\leq4$.

Then, in the eigenbasis of $\bar C$, the matrix $C_N$ is diagonal with eigenvalues $N_1\lambda_i+2N_2$. Defining the fraction $t=N_1/N$ of the sample size used in stage (1), we obtain
\begin{align}
    \bar I\lesssim f(t)\,\frac{d-1}{N}, \label{eq:mean_f}
\end{align}
with
\begin{align}
    f(t)=\frac{1}{d-1}\sum_{i=1}^{2(d-1)}\frac{1}{t\lambda_i+2(1-t)}. \label{eq:f_of_t_SM}
\end{align}
Since $\lambda_i\geq0$, each term of the sum is at most $1/[2(1-t)]$, so that $f(t)\leq1/(1-t)$. This value corresponds to discarding the counts of stage (1). Moreover, since the function $1/[tx+2(1-t)]$ is convex and the average of the $\lambda_i$ is $2$, Jensen's inequality gives $f(t)\geq1$, with equality only when $\bar C$ is isotropic. We can see that the average infidelity lies between the Gill-Massar bound $(d-1)/N$ and $(d-1)/N_2$.

\subsection{Bound for the eigenvalues of the Fisher Information matrix}\label{sec:fim_stage1}

The eigenvalues $\lambda_i$ of the Fisher information matrix $\bar C$ of stage (1) depend on the unknown state, which is in general far from the fiducial state of $\mathcal{E}_\pm$, so that we cannot calculate them analytically. Here, we find a lower bound for them. The strategy is the following. First, we obtain the probabilities of the outcomes of $\mathcal{E}_\pm$. Then, we calculate their derivatives with respect to the magnitude and the phase of each coefficient of the state. Finally, we use these derivatives to obtain the Fisher information along these directions, which gives the eigenvalues of $\bar C$.

In order to obtain the probabilities, we expand the unknown state in the basis in which the fiducial state of $\mathcal{E}_\pm$ is $\ket*{0}$, that is, $\ket*{\Psi}=a_0\ket*{0}+\sum_{k=1}^{d-1}z_k\ket*{k}$, where the coefficients $z_k=x_{k0}+ix_{k1}$ are defined below Eq.\thinspace\eqref{eq:probs_2fsm} and are not necessarily small. For the explicit FSMs of section \ref{sec:exp_cons_fsm}, the elements $2k-1$ and $2k$ contain the terms $z\ket*{k}$ and $z^*\ket*{k}$, where $z$ is a constant. Using $e^{-i\pi/4}z^*=e^{-i\pi/4}/(\sqrt n+1)-\sqrt{n/2}$, their amplitudes are
\begin{align}
    \braket*{\varphi^{2k-1}_\pm}{\Psi}=\frac{1}{\sqrt n}\left(A_\pm\mp\sqrt{\frac{n}{2}}\,z_k\right),
\end{align}
and
\begin{align}
    \braket*{\varphi^{2k}_\pm}{\Psi}=\frac{1}{\sqrt n}\left(A_\pm\pm i\sqrt{\frac{n}{2}}\,z_k\right),
\end{align}
where the signs correspond to $\mathcal{E}_\pm$ and
\begin{align}
    A_\pm=a_0\pm\frac{e^{-i\pi/4}}{\sqrt n+1}\sum_{j=1}^{d-1}z_j.
\end{align}
That is, each pair of outcomes measures one coefficient $z_k$ through the interference with $A_\pm$. Then, the probabilities are
\begin{align}
    p_{2k-1}=\frac{1}{n}\left[|A_\pm|^2+\frac{n}{2}|z_k|^2\mp\sqrt{2n}\,{\rm Re}(A_\pm^*z_k)\right],
\end{align}
and
\begin{align}
    p_{2k}=\frac{1}{n}\left[|A_\pm|^2+\frac{n}{2}|z_k|^2\pm\sqrt{2n}\,{\rm Re}(iA_\pm^*z_k)\right].
\end{align}
For a typical state, $(n/2)|z_k|^2\approx1$ and $|A_\pm|\approx1/\sqrt d$. Thereby, both probabilities are approximately $|z_k|^2/2$. We notice that the magnitude of $z_k$ appears in the dominant term, while its phase appears only in the interference terms.

Now we calculate the derivatives of the probabilities with respect to a normalized displacement $\varepsilon$ of the state. If the displacement changes $|z_k|$, it changes the dominant term, and the derivatives of both probabilities are approximately $|z_k|$. Dividing their squares by the probabilities, the Fisher information along this direction is approximately $4$. If the displacement changes the phase of $z_k$, the derivatives are $\mp\sqrt{2/n}\,{\rm Re}(iA_\pm^*z_k)/|z_k|$ and $\mp\sqrt{2/n}\,{\rm Re}(A_\pm^*z_k)/|z_k|$. The sum of their squares is $2|A_\pm|^2/n$, and dividing by the probabilities we obtain the Fisher information $4|A_\pm|^2/(n|z_k|^2)$. Averaging over $\mathcal{E}_\pm$, the cross terms of $|A_\pm|^2$ cancel, and we have
\begin{align}
    |A|^2=a_0^2+\frac{1}{(\sqrt n+1)^2}\left|\sum_{j=1}^{d-1}z_j\right|^2\geq a_0^2.
\end{align}
Since the phases of different coefficients are coupled only through $\sum_jz_j/(\sqrt n+1)$, to leading order in $1/d$ the phase directions are eigenvectors of $\bar C$ with the smallest eigenvalues
\begin{align}
    \lambda_k\approx\frac{4|A|^2}{n|z_k|^2}\geq\frac{4}{n}\,\frac{a_0^2}{|z_k|^2},\qquad k=1,\dots,d-1. \label{eq:lmin_ratio}
\end{align}
Finally, stage (0) selects a component whose overlap is comparable to the largest one, see section~\ref{sec:afsm}, so that the ratio $a_0^2/\max_k|z_k|^2$ in Eq.~\eqref{eq:lmin_ratio} is of order one. When the selected component is the largest one, this ratio is at least one, and all the eigenvalues of $\bar C$ satisfy
\begin{align}
    \lambda_i\gtrsim\frac{4}{n}=\frac{4}{2d-1}. \label{eq:lmin_theory}
\end{align}

\subsection{High-probability bounds}\label{sec:mle_hp}

In order to bound the infidelity of a single estimate, we use the following concentration inequality \cite{Laurent2000}. If $g_1,\dots,g_{2(d-1)}$ are independent Gaussian random variables with zero mean and variances $\nu_i$, then for any $x>0$
\begin{align}
    \mathbb{P}\left[\sum_{i}g_i^2\geq\sum_{i}\nu_i+2\sqrt{x\sum_{i}\nu_i^2}+2x\max_i\nu_i\right]\leq e^{-x}. \label{eq:LM}
\end{align}
For equal variances, Eq.\thinspace\eqref{eq:LM} reduces to the usual bound on the quantile of the $\chi^2$ distribution. Here, the variances are different, since the counts of stage (1) contribute anisotropically to the Fisher information. We apply Eq.\thinspace\eqref{eq:LM} to the components of the error along the eigenvectors of the covariance matrix, whose squares add up to $|\bm x-\bm y|^2$.

We first apply this inequality to the estimator of stage (1). The first estimate is the fiducial state of $\tilde{\mathcal{E}}$, and its infidelity is $1-\bar a_0^2$, with $\bar a_0$ the overlap defined in section \ref{sec:afsm}. The MLE of stage (1) has covariance $(N_1\bar C)^{-1}$, whose eigenvalues are $1/(N_1\lambda_i)$. Choosing $x=\log(1/\delta_1)$ in Eq.\thinspace\eqref{eq:LM}, we obtain, with probability approaching $1-\delta_1$,
\begin{align}
    1-\bar a_0^2\lesssim\frac{1}{N_1}\left(\sum_{i}\frac{1}{\lambda_i}+2\sqrt{\log\frac{1}{\delta_1}\sum_{i}\frac{1}{\lambda_i^2}}+\frac{2}{\lambda_{\min}}\log\frac{1}{\delta_1}\right), \label{eq:eps1_MLE}
\end{align}
where $\lambda_{\min}$ is the smallest eigenvalue of $\bar C$. Then, we bound each $1/\lambda_i$ by $1/\lambda_{\min}$. According to section \ref{sec:fim_stage1}, the smallest eigenvalue satisfies $\lambda_{\min}\gtrsim4/(2d-1)$, Eq.\thinspace\eqref{eq:lmin_theory}.
Replacing Eq.\thinspace\eqref{eq:lmin_theory} in Eq.\thinspace\eqref{eq:eps1_MLE}, we obtain
\begin{align}
    1-\bar a_0^2\lesssim\frac{(d-1)(2d-1)}{2N_1}\left[1+\sqrt{\frac{2\log(1/\delta_1)}{d-1}}+\frac{\log(1/\delta_1)}{d-1}\right]. \label{eq:s1_d2}
\end{align}
We notice that the infidelity of the first estimate scales as $d^2/N_1$, while the other terms in brackets vanish for large $d$.

Then, we apply the concentration inequality to the final estimate, which combines the data of stages (1) and (2) and has covariance $C_N^{-1}$. In order to apply Eq.\thinspace\eqref{eq:LM}, we need the largest eigenvalue $\max_i\nu_i$ of $C_N^{-1}$ and the sum $\sum_i\nu_i^2$. Since $\bar C$ has non-negative eigenvalues, Eq.\thinspace\eqref{eq:CN} gives $\bm v^\top C_N\bm v\geq2N_2$ for any normalized vector $\bm v$. Thereby, $\max_i\nu_i\leq1/(2N_2)$ and $\sum_i\nu_i^2\leq(d-1)/(2N_2^2)$. Choosing $x=\log(1/\delta_2)$ and using Eq.\thinspace\eqref{eq:mean_f}, we obtain, with probability approaching $1-\delta_2$,
\begin{align}
    I(\Psi(\bm x),\Psi(\bm y))\lesssim f(t)\,\frac{d-1}{N}+\frac{1}{N_2}\left(\sqrt{2(d-1)\log\frac{1}{\delta_2}}+\log\frac{1}{\delta_2}\right). \label{eq:MLE_bound}
\end{align}
The first term is the average infidelity. The second term describes the fluctuations of a single estimate around the average, and it scales as $\mathcal{O}(\sqrt d/N)$.

\subsection{Optimal splitting of the sample size}

In order to find the optimal fraction $t$, we expand $f(t)$ for small $t$. Each term of Eq.\thinspace\eqref{eq:f_of_t_SM} can be written as $1/[2+t(\lambda_i-2)]$. Expanding to second order in $t$, the first-order terms cancel because the average of the $\lambda_i$ is $2$, and we obtain
\begin{align}
    f(t)=1+\frac{t^2}{8(d-1)}\sum_{i}(\lambda_i-2)^2+\mathcal{O}(t^3). \label{eq:f_small_t}
\end{align}
We can see that the samples of stage (1) do not increase the infidelity to first order in $t$. Since $0\leq\lambda_i\leq4$, we have $(\lambda_i-2)^2\leq4$, and the coefficient of $t^2$ is at most $1$. Thereby, $f(t)\lesssim1+t^2$.

Eq.\thinspace\eqref{eq:f_small_t} assumes that $\tilde C=2\mathbb{I}$. However, for a finite $N_1$, the first estimate differs from $\ket*{\Psi}$, and the eigenvalues of $\tilde C$ become $2+\mu_j$, where $\mu_j$ are the deviations of the eigenvalues from $2$, with $\sum_j\mu_j=0$ according to Eq.\thinspace\eqref{eq:trC}. For small $t$, the infidelity is dominated by stage (2), and expanding ${\rm tr}(N_2\tilde C)^{-1}$ to second order in $\mu_j$ we obtain
\begin{align}
    \bar I\approx\frac{d-1}{N_2}+\frac{1}{8N_2}\sum_j\mu_j^2. \label{eq:mismatch}
\end{align}
We notice that the mismatch increases the infidelity only to second order in $\mu_j$. Evaluating numerically the Fisher information of an FSM whose fiducial state is displaced from $\ket*{\Psi}$, and averaging over the direction of the displacement, we find $\sum_j\mu_j^2\approx8(d-1)(1-\bar a_0^2)$. Then, the mismatch adds the infidelity $1-\bar a_0^2$ of the first estimate to $f(t)$. Using Eq.\thinspace\eqref{eq:s1_d2} to leading order in $d$ and $N_1=tN$, we obtain, with probability approaching $1-\delta_1$,
\begin{align}
    f(t)\lesssim1+t^2+\frac{(d-1)(2d-1)}{2tN}. \label{eq:f_expansion}
\end{align}
The quadratic term increases with $t$, since stage (1) is less informative than stage (2). The last term decreases with $t$, since the first estimate becomes more accurate. The right-hand side of Eq.\thinspace\eqref{eq:f_expansion} is minimal at
\begin{align}
    t^*=\left[\frac{(d-1)(2d-1)}{4N}\right]^{1/3}, \label{eq:optimal_split_SM}
\end{align}
which gives the sample sizes
\begin{align}
    N_1^*=\left[\frac{(d-1)(2d-1)}{4}\right]^{1/3}N^{2/3}, \label{eq:N1_opt_SM}
\end{align}
and $N_2^*=N-N_1^*$. We can see that the fraction of samples used in stage (1) decreases as $N^{-1/3}$. At the optimum, the last term of Eq.\thinspace\eqref{eq:f_expansion} is twice the quadratic term, and $t^{*2}=(d-1)(2d-1)/(4N_1^*)$. Replacing these results in Eq.\thinspace\eqref{eq:MLE_bound}, and keeping the term in brackets of Eq.\thinspace\eqref{eq:s1_d2}, we obtain, with probability approaching at least $1-\delta_1-\delta_2$,
\begin{align}
    I(\Psi(\bm x),\Psi(\bm y))\lesssim\;&\frac{d-1}{N}\left\{1+\frac{(d-1)(2d-1)}{2N_1^*}\left[\frac{3}{2}+\sqrt{\frac{2\log(1/\delta_1)}{d-1}}+\frac{\log(1/\delta_1)}{d-1}\right]\right\}\nonumber\\
    &+\frac{1}{N_2^*}\left(\sqrt{2(d-1)\log\frac{1}{\delta_2}}+\log\frac{1}{\delta_2}\right). \label{eq:MLE_bound_opt}
\end{align}
The first term is the Gill-Massar bound multiplied by a correction due to stage (1), which decreases as $1/N_1^*$. This correction multiplies the Gill-Massar bound because the first estimate affects the final estimate only through the Fisher information of $\tilde{\mathcal{E}}$. Replacing $N_1^*$ in Eq.\thinspace\eqref{eq:N1_opt_SM}, we can see that the correction is of the order of $(d^2/N)^{2/3}$. Thereby, for $N\gg d^2$ the correction vanishes, and the first term approaches the Gill-Massar bound, that is, the optimal precision. The second term describes the fluctuations of a single estimate of stage (2) around the average, and it decreases as $1/N_2^*$, that is, it is of the order of $\sqrt d/N$. Therefore, for $N\gg d^2$, the average infidelity of the MLE approaches the optimal value given by the Gill-Massar bound and a single estimate deviates from it by at most $\mathcal{O}(\sqrt d/N)$. Another advantage of the MLE is that it does not exhibit the bias $\kappa$ of the linear estimators, since it uses the exact probabilities and the combined data of $\mathcal{E}_\pm$ and $\tilde{\mathcal{E}}$ are informationally complete for any pure state with $a_0\neq0$.

%%%%%%%%%%%%%%%%%%%%%%%%%%%%%%%%%%%%%%%%%%%%%%%%%%%%%%%%%%%%%%%%%%%%%%%
%-------------------------------------------------------------------------
    \section{Estimation of mixed states}
    In this section, we study the estimation of slightly mixed states, pure states affected by white noise. They consist of a pure state component plus a maximally mixed state,
    \begin{equation}
    \rho(\bm{x}, \lambda) = (1-\lambda)\ket*{\Psi}\bra*{\Psi} + \frac{\lambda}{d}\mathbb{I}. 
    \end{equation}
    where $\lambda \in [0,1]$ is an unknown constant and $\bm{x}=[x_{0,0},x_{0,1},\cdots, x_{d-1,0}, x_{d-1,1}]^\top$ are real parameters, which are not assumed to be small. Considering the form of the pure state on the computational basis,
    \begin{equation}
    \ket*{\Psi} = a_0\ket*{0} + \sum_{k=1}^{d-1}(x_{k0} + i x_{k1})\ket*{k}.
    \end{equation}
    We will demonstrate that measurements into two FSMs $\mathcal{E}_\pm = \{ \ket*{\varphi_\pm^\alpha}\bra*{\varphi_\pm^\alpha}\}_{\alpha=1,\dots,n}$ and the computational basis $\{\ket*{k} \}_{k=0}^{d-1}$ are enough to characterize the pure state $\ket*{\Psi}$ and the parameter $\lambda$.

    \subsection{Analytical reconstruction}

    The probabilities of measuring the state $\rho$ on the computational basis $\{\ket*{k}\}_{k=0}^{d-1}$ are
    \begin{align}
        P_k &= \bra*{k}\rho \ket*{k}, \\
        &= (1-\lambda)|\langle k|\Psi\rangle|^2 + \frac{\lambda}{d}, \\
        &= (1-\lambda)|z_k|^2 + \frac{\lambda}{d},
    \end{align}
    with $z_k=x_{k0}+ix_{k1}$ and $z_0=a_0$ for simplicity. Rearranging terms, we find a set of equations for the coefficients $z_k$,
    \begin{equation}
        |z_k|^2 = \frac{P_k - \lambda/d}{1 - \lambda}.
        \label{probabilidad base computacional}
    \end{equation}
    We chose the fiducial state with the highest probability of measurement.
    
    Considering the FSM with elements,
    \begin{equation}
    \ket*{\varphi_\pm ^\alpha} = \beta_0^\alpha \ket*{0} \pm \sum_{k=1}^{d-1} \omega_k^\alpha\ket*{k}, \; \; \beta_0^\alpha>0,  \; \alpha=0,\dots,n-1,
    \end{equation}
    where  $\omega_k^\alpha = \beta_k^\alpha + i \gamma_k^\alpha$, and $n\geq 2d-1$ the number of rank-1 elements of the FSM. And fiducial state the one chosen in the previous step. The probabilities of measuring the slightly mixed state $\rho$ with the FSM are given by
    \begin{align}
    P_\pm^\alpha &= \Tr(\mathcal{E}_\pm^\alpha \; \rho ) , \\
    &= (1-\lambda)|\langle\varphi_\pm^\alpha|{\Psi}\rangle|^2 + \frac{\lambda}{d} \langle\varphi_\pm^\alpha|\varphi_\pm^\alpha\rangle.
    \end{align}
    By calculating the projection of the pure state onto the measurement elements, we obtain
    \begin{align}
    | \braket*{\varphi_\pm^\alpha}{\Psi} |^2 = (\beta_0^\alpha)^2 a_0^2 + \left|\sum_{j=1}^{d-1} (\omega_j^\alpha)^* z_j \right|^2 \pm \beta_0^\alpha a_0\sum_{j=1}^{d-1}\left[\left(\omega_j^\alpha\right)^* z_j + \omega_j^\alpha z_j^* \right].
    \end{align}
    In addition, defining
    \begin{align}
    \varphi^\alpha &= \langle\varphi_\pm^\alpha | \varphi_\pm^\alpha \rangle\\
    &= \left(\beta_0^\alpha \langle 0| \pm \sum_{k=1}^{d-1}(\omega_k^\alpha)^*\langle k|\right)\left(\beta_0^\alpha |0\rangle \pm \sum_{k=1}^{d-1}\omega_k^\alpha |k\rangle\right), \\
    &= (\beta_0^\alpha)^2 + \sum_{k=1}^{d-1}|\omega_k^\alpha|^2,
    \end{align}
    the probabilities become
    \begin{align}
    P_\pm^\alpha &= (1-\lambda)\left[ (\beta_0^\alpha)^2 a_0^2 + \left|\sum_{j=1}^{d-1} (\omega_j^\alpha)^* z_j \right|^2 \pm \beta_0^\alpha a_0\sum_{j=1}^{d-1}\left[\left(\omega_j^\alpha\right)^* z_j + \omega_j^\alpha z_j^* \right] \right] + \frac{\lambda}{d} \varphi^\alpha .
    \end{align}
    Now, by adding and subtracting these expressions, we obtain
    \begin{align}
    P_+^\alpha - P_-^\alpha &= 2(1-\lambda)\beta_0^\alpha a_0 \sum_{j=1}^{d-1}\left[\left(\omega_j^\alpha\right)^* z_j + \omega_j^\alpha z_j^* \right], \label{subtraction}\\
    P_+^\alpha + P_-^\alpha &= 2(1-\lambda)\left[ (\beta_0^\alpha a_0)^2 + \left|\sum_{j=1}^{d-1} (\omega_j^\alpha)^* z_j \right|^2 \right] + 2 \frac{\lambda}{d} \varphi^\alpha. \label{addition}
    \end{align}
    From the subtraction of the probabilities \eqref{subtraction}, we can define the following expression
    \begin{align}
    \Delta_k &= \frac{1}{2}\sum_{\alpha=1}^n \left(\frac{\omega_k^\alpha}{\beta_0^\alpha}\right)(P_+^\alpha - P_-^\alpha)  , \\
    &= (1-\lambda)a_0 z_k. 
    \label{Deltak}
    \end{align}
    To isolate the parameters $z_k$, we have to find the parameter $a_0$. This can be done by noticing that
    \begin{equation}
    |\Delta_k|^2 = (1-\lambda)^2 a_0^2 \,|z_k|^2. 
    \label{delta k cuadrado}
    \end{equation}
    From the probabilities in the computational basis, we obtain the expressions \eqref{probabilidad base computacional} for the coefficients $|z_k|^2$ as a function of $\lambda$. Substituting into the previous equation,
    \begin{equation}
    |\Delta_k|^2 = (P_0 - \lambda/d)(P_k - \lambda/d) , \; k=1,\dots,d-1,
    \end{equation}
    we arrive at a quadratic equation for $\lambda$,
    \begin{equation}
    \lambda^2 - \lambda \, d(P_0 + P_k) + d^2\, (P_0P_k - |\Delta_k|^2) = 0.
    \end{equation}
    With solutions,
    \begin{equation}
    \lambda_k = \frac{d}{2}\left[P_0 + P_k \pm \sqrt{(P_0-P_k)^2 + 4|\Delta_k|^2}\right].
    \end{equation}
    Since $\lambda$ is small, we consider the negative solution. As we have $d-1$ equations for $\lambda$, we take the average of the solutions as our estimate, keeping the negative values produced by finite statistics and clipping the result at zero,
    \begin{equation}
    \lambda = \frac{1}{d-1}\sum_{k=1}^{d-1}\lambda_k.
    \end{equation}
    The purity of the state $\rho$ is related to the value of $\lambda$ by
    \begin{equation}
    \Tr(\rho^2) = (1-\lambda)^2 + \frac{\lambda(2-\lambda)}{d}.
    \end{equation}
    This demonstrates that the tomography protocol enables us to verify the purity of the state. Knowing $\lambda$, we can obtain estimates of the coefficients $a_0$ and $z_k$ from equation \eqref{probabilidad base computacional}. However, we can also obtain them from the following quadratic equation,
    \begin{equation}
    \sum_{k=1}^{d-1} |\Delta_k|^2 = (1-\lambda)^2 a_0^2 (1-a_0^2),
    \end{equation}
    From which we obtain a quadratic equation for $a_0^2$,
    \begin{equation}
    a_0^4 - a_0^2 + \frac{1}{(1-\lambda)^2}\sum_{k=1}^{d-1}|\Delta_k|^2 = 0.
    \label{cuadratica1}
    \end{equation}
    Now, from the expression for the sum of probabilities, we have that
    \begin{align}
    P_+^\alpha + P_-^\alpha &= 2(1-\lambda)\left[ (\beta_0^\alpha a_0)^2 + \left|\sum_{k=1}^{d-1}(\omega_k^\alpha)^* z_k \right|^2\right] + 2 \frac{\lambda}{d} \varphi^\alpha.
    \end{align}
    Furthermore, from the expression for the difference of probabilities, we can rearrange to obtain
    \begin{equation}
     z_k = \frac{\Delta_k}{a_0(1-\lambda)}.
    \end{equation}
    Substituting this into the previous expression, we get
    \begin{align}
    P_+^\alpha + P_-^\alpha &= 2(1-\lambda)\left[ (\beta_0^\alpha a_0)^2 + \left|\sum_{k=1}^{d-1}(\omega_k^\alpha)^* \frac{\Delta_k}{a_0(1-\lambda)}\right|^2\right] + 2\frac{\lambda}{d} \varphi^\alpha.
    \end{align}
    Rearranging terms, we obtain an equation for $a_0$,
    \begin{equation}
    a_0^4(1-\lambda)^2(\beta_0^\alpha)^2 - a_0^2(1-\lambda)((P_+^\alpha + P_-^\alpha)/2 - \frac{\lambda}{d}\varphi^\alpha) + \left|\sum_{k=1}^{d-1}(\omega_k^\alpha)^*\Delta_k\right|^2 = 0.
    \label{cuadratica2}
    \end{equation}

    By equating equations \eqref{cuadratica1} and \eqref{cuadratica2}, we obtain a quadratic equation for $a_0$,
    \begin{align}
    a_0^2 - \frac{1}{(1-\lambda)^2}\sum_{k=1}^{d-1}|\Delta_k|^2 &=  a_0^2\frac{((P_+^\alpha + P_-^\alpha)/2 - \frac{\lambda}{d}\varphi^\alpha)}{(1-\lambda)(\beta_0^\alpha)^2} - \frac{|\sum_{k=1}^{d-1}(\omega_k^\alpha)^* \Delta_k|^2}{(1-\lambda)^2(\beta_0^\alpha)^2}, \\
    \end{align}
    Since $a_0>0$, we consider the positive solution,
    \begin{equation}
    a_0^\alpha = \sqrt{\frac{ |\sum_{k=1}^{d-1}(\omega_k^\alpha)^*\Delta_k|^2 - (\beta_0^\alpha)^2\sum_{k=1}^{d-1}|\Delta_k|^2 }{(1-\lambda)\left[(P_+^\alpha + P_-^\alpha)/2 - (1-\lambda)(\beta_0^\alpha)^2 - \frac{\lambda}{d}\varphi^\alpha\right]}}.
    \end{equation}
    Noting that we have a solution for each $\alpha$, we take the median of the solutions with $0<(a_0^\alpha)^2<1$ as our estimate for $a_0$,
    \begin{equation}
    a_0 = {\rm median}\{a_0^\alpha\}.
    \end{equation}
    Finally, having the estimated values of $\lambda$ and $a_0$, we can find the values of the coefficients $z_k$ from expression \eqref{Deltak},
    \begin{equation}
    z_k = \frac{\Delta_k}{(1-\lambda)a_0}.
    \end{equation}
    Thereby, we can estimate the unknown slightly mixed state $\rho$. \\
    
    Given that $\rho$ is highly pure, we expect that performing a near-optimal FSM for the state $\ket*{\Psi}$ will improve the accuracy of its estimation. To achieve this, we use the preliminary estimator obtained from measurements in the computational basis together with the two non-optimal FSMs $\mathcal{E}_\pm$ to construct a near-optimal FSM $\mathcal{E}$ for $\ket*{\Psi}$. By maximizing the likelihood of all counts of these settings, where each measurement contributes with its number of counts, we obtain an enhanced final estimator of $\rho$.

\subsection{Error bound and optimal splitting}\label{sec:mixed_split}
The optimal splitting of section \ref{sec:mle} also holds for slightly mixed states. With white noise, the probabilities of any POVM become $(1-\lambda)p_\alpha+\lambda\,{\rm Tr}(E^\alpha)/d$, where $p_\alpha$ are the probabilities of the pure state $\ket*{\Psi}$. For the adapted FSM, all the probabilities at the fiducial state are equal, so that the noise does not change them, while their derivatives with respect to the parameters of $\ket*{\Psi}$ are multiplied by $1-\lambda$. Thereby, the Fisher information of stage (2) is multiplied by $(1-\lambda)^2$. For the FSMs of stage (1), the noise mainly affects the outcomes with small probabilities, and their Fisher information is approximately multiplied by the same factor. Then, the infidelities in section \ref{sec:mle} are divided by approximately $(1-\lambda)^2$. In particular, the error bound of Eq.\thinspace\eqref{eq:MLE_bound_opt} also holds for slightly mixed states, with its right-hand side divided by approximately $(1-\lambda)^2$. Moreover, the optimal fraction in Eq.\thinspace\eqref{eq:optimal_split_SM} is multiplied only by $(1-\lambda)^{-2/3}$, which is close to one for slightly mixed states. Therefore, we also use the sample size $N_1^*$ from Eq.\thinspace\eqref{eq:N1_opt_SM} for mixed states. Let us note that the noise parameter $\lambda$ must also be estimated, and that for $\lambda>0$ the FSMs do not saturate the Gill-Massar inequality. Thereby, the pure-state Gill-Massar bound is only a reference for mixed states.

\end{document}